\documentclass[12pt]{article}%
\pdfoutput=1

\usepackage{hyperref}
\usepackage[nosort]{cite}
\usepackage{graphicx}
\usepackage{xypic}
\usepackage{subcaption}
\usepackage{multicol}
\usepackage{wrapfig}
\usepackage{amsfonts}
\usepackage{amssymb,lipsum}
\usepackage{amsmath}
\usepackage{heck}
\usepackage{setspace}
\usepackage{verbatim}
\usepackage{color}
\usepackage{longtable}
\usepackage{float}
\usepackage{epstopdf}
\usepackage{tikz}
\usepackage[margin=1in]{geometry}
\usepackage{titletoc}%
\providecommand{\U}[1]{\protect\rule{.1in}{.1in}}
\pdfoutput=1
\usetikzlibrary{decorations.markings}

\numberwithin{equation}{section}

\newcommand{\ba}{\begin{eqnarray}}
\newcommand{\ea}{\end{eqnarray}}

\newcommand{\cI}{\mathcal{I}}

 \newcommand{\PP}{\mathbb{P}^1}
 \newcommand{\CE}{\mathcal{E}}

 \newtheorem{theorem}{Theorem}[section]

\newtheorem{definition}[theorem]{Definition}
\newtheorem{remark}[theorem]{Remark}

\newcommand{\beq}{\begin{equation}}
\newcommand{\eeq}{\end{equation}}
\newcommand{\bea}{\begin{equation}\begin{aligned}}	
\newcommand{\eea}{\end{aligned}\end{equation}}		
\begin{document}

\date{February 2017}

\title{T-Branes at the Limits of Geometry}

\institution{VTECH}{\centerline{${}^{1}$Physics Department, Robeson Hall, Virginia Tech, Blacksburg, VA 24061, USA}}

\institution{UNC}{\centerline{${}^{2}$Department of Physics, University of North Carolina, Chapel Hill, NC 27599, USA}}

\institution{UIUC}{\centerline{${}^{3}$Department of Mathematics, University of Illinois at Urbana-Champaign, Urbana, IL 61801, USA}}

\institution{UIC}{\centerline{${}^{4}$Department of Mathematics, University of Illinois at Chicago, Chicago, IL 60607, USA}}

\authors{Lara B. Anderson\worksat{\VTECH}\footnote{e-mail: {\tt lara.anderson@vt.edu}},
Jonathan J. Heckman\worksat{\UNC}\footnote{e-mail: {\tt jheckman@email.unc.edu}},\\[4mm]
Sheldon Katz\worksat{\UIUC}\footnote{e-mail: {\tt katzs@illinois.edu}} and
Laura P. Schaposnik\worksat{\UIC}\footnote{e-mail: {\tt schapos@uic.edu}}
}

\abstract{Singular limits of 6D F-theory compactifications
are often captured by T-branes, namely a non-abelian configuration of
intersecting 7-branes with a nilpotent matrix of normal deformations.
The long distance approximation of such 7-branes is
a Hitchin-like system in which simple and irregular poles
emerge at marked points of the geometry. When multiple matter fields localize
at the same point in the geometry, the associated Higgs field
can exhibit irregular behavior, namely poles of
order greater than one. This provides a
geometric mechanism to engineer wild Higgs bundles.
Physical constraints such as anomaly cancellation and consistent coupling to gravity
also limit the order of such poles. Using this geometric formulation, we  unify seemingly different
wild Hitchin systems in a single framework in which orders of poles become adjustable parameters dictated by
tuning gauge singlet moduli of the F-theory model.}

\enlargethispage{\baselineskip}

\setcounter{tocdepth}{2}

\renewcommand\Large{\fontsize{15}{17}\selectfont}

\maketitle

\tableofcontents

\newpage

\section{Introduction \label{sec:INTRO}}

There is a close interplay between the geometry of extra dimensions
in string theory and low energy effective field theory.
In a theory of open and closed strings
it is common to associate geometry with closed string modes such as the graviton,
and field theory sectors with open string modes. This leads to a
physically rich space of vacua.

An important goal in string compactification is to characterize all resulting
effective field  theories. One general lesson is that the open and closed string sectors often
provide complementary pictures, much as one would assign coordinate patches
on a manifold. Celebrated examples include open/closed string channel
duality, and the AdS/CFT correspondence \cite{juanAdS}. A priori, however, there is
no reason to expect a single patch to cover all regimes. From this perspective,
the important question is to determine the transition functions required to move from one patch to another.

This is particularly pressing in F-theory, where the backreaction of 7-branes on the 10D spacetime is encoded in terms of an auxiliary 12D geometry given by a torus fibration over the 10D spacetime. This ``closed string'' perspective is quite helpful in determining how to consistently couple 7-branes to gravity. In this approach, one also encounters singular regions in the 10D spacetime where the torus fibration degenerates. In such situations, the geometric picture breaks down, and one instead passes to the gauge theory on a 7-brane, namely the open string sector.

But in general the moduli space of the 7-brane gauge theory will contain more
than just classical commutative geometry. This is
because the degrees of freedom for 7-branes are captured by matrix degrees
of freedom. As such, typically only the eigenvalues of a matrix translate into
commutative geometry. When the matrix degrees of freedom do not commute, we
pass to a more general configuration in which the 7-brane puffs up in directions transverse to its worldvolume. This is known as a T-brane \cite{Donagi:2003hh, TBRANES},
as the matrix of normal deformations is upper triangular. For recent work
on the formal structure of T-branes in F-theory, see e.g. \cite{Anderson:2013rka, Collinucci:2014qfa,
Collinucci:2014taa, Collinucci:2016hpz, Bena:2016oqr}.
For phenomenological applications of T-branes,
see e.g. \cite{TBRANES, Chiou:2011js, Font:2012wq, Font:2013ida, Marchesano:2015dfa, Cicoli:2015ylx,Marchesano:2016cqg,Ashfaque:2017iog}.
For reviews on F-theory model building, see e.g. \cite{HVLHC, Heckman:2010bq, Weigand:2010wm,
Maharana:2012tu, Wijnholt:2012fx}.

In this paper we study T-brane vacua for 6D and 4D theories with eight real supercharges.
More precisely, we consider F-theory compactified on an elliptically fibered Calabi-Yau threefold. This yields a
6D theory with $\mathcal{N} = (1,0)$ supersymmetry. Further compactification on a $T^2$ yields a 4D $\mathcal{N} = 2$
theory which we can alternatively study using type IIA string theory compactified on the same Calabi-Yau threefold.
Our goal will be to take steps towards a general prescription for the limiting behavior of T-branes as we pass
from the ``open string patch'' of moduli space to the ``closed string patch'' captured by Calabi-Yau geometry.

This point of view can lead to a sharp correspondence between the moduli spaces of a Hitchin
system \cite{HitchinSelf} on a Riemann surface $C$ with gauge group $G$ of ADE type, and the
local Calabi-Yau threefold $X$ associated with a curve of ADE
singularities \cite{Anderson:2013rka, Diaconescu:2005jw, Diaconescu:2006ry}.
Recall that in the Hitchin system we have an adjoint valued $(1,0)$ form
$\Phi$ and a gauge connection $A$. Gauge invariant Casimir invariants
of $\Phi$ translate in the type IIA Calabi-Yau geometry to complex structure deformations, while periods of the gauge field (more precisely its holonomies) translate to periods of the Ramond-Ramond three-form potential, with values in the intermediate Jacobian $H^3(X,\mathbb{R})/H^3(X,\mathbb{Z})$. The transition between these two descriptions of patches
of moduli space is captured by the theory of limiting mixed Hodge structures \cite{Anderson:2013rka}. There a global/compact description of T-branes was proposed that described the Hitchin moduli space (open string degrees of freedom) as ``emergent" in a singular limit of the CY geometry. The precise correspondence between Hitchin and singular CY moduli was laid out in \cite{Anderson:2013rka} and can be summarized by the following diagram

\beq\label{emergent_hitchin}
\xymatrix@R=15pt@C=15pt{
& M \ar[d] \\
\pi^* H \ar[ur] \ar[r] \ar[d] & \widetilde M_{\mathrm{cplx}} \ar[d]^\pi \\
H \ar[r] & M_{\mathrm{loc}}
}
\eeq
where $H$ and $M$ are the full Hitchin and Calabi-Yau moduli spaces, respectively, and $\widetilde{M}_{\mathrm{cplx}}$ and $M_{loc}$ the complex
structure moduli spaces of the resolved Calabi-Yau geometry and local (singularity preserving) complex structure deformations of the singular Calabi-Yau geometry (note: the bottom map is the Hitchin fibration and the top map is an inclusion).

This prompts a number of natural questions:
\begin{itemize}
\item Does this correspondence extend to singular field configurations of the Hitchin system? And in what singular Calabi-Yau geometries might these arise?

\item Does F-theory impose physical constraints on such singularities?
\end{itemize}
While we will not fully resolve these questions in the present work, our aim in this paper will be to show that to a large extent, there is a natural extension to the case of
Higgs fields with singularities, which again can lead to a perfect match between open and closed string moduli.
In addition, we will also show that the overall type of singularities
which can be engineered are often constrained by the further condition that a compact
F-theory or type IIA background really exists.

In physical terms, the Hitchin system on a Riemann surface $C$ emerges
as the long distance description of 7-branes
wrapped on $C$. It is specified by introducing a gauge group $G$,
an adjoint valued $(1,0)$-form $\Phi$, and a gauge field $A$.
Solutions to the equations of motion at generic points of $C$ are \cite{HitchinSelf}:
\begin{equation}
\overline{\partial}_A \Phi = 0 \text{\,\,\,and\,\,\,} F + [\Phi,\Phi^\dag] = 0,
\end{equation}
modulo gauge transformations.
The correspondence between the moduli of this system and the associated
local curve of ADE singularities has been studied in references
\cite{Anderson:2013rka, Diaconescu:2005jw, Diaconescu:2006ry, BHVI}.
T-Branes correspond to the special class of configurations where $\Phi$
is nilpotent in the Lie algebra. For a matrix valued $\Phi$, i.e., for the classical
algebras, this amounts to the condition $\mathrm{Tr}(\Phi^l) = 0$ for sufficiently large $l$.
As the moduli space of the Hitchin system (with smooth Higgs field) is connected, there is a sense in which we can
build up quite general solutions starting from a T-brane configuration. Indeed, starting from such a configuration,
we can perform perturbations in the entries of this solution, thus realizing a broad class of additional
solutions \cite{HitchTeich}. In the dual frame of heterotic string constructions, this is the statement that there is a
single connected component to the moduli space of stable holomorphic vector bundles on a $K3$ surface.

\begin{figure}[t!]%
\centering
\includegraphics[
scale = 0.50, trim = 15mm 30mm 0mm 30mm]%
{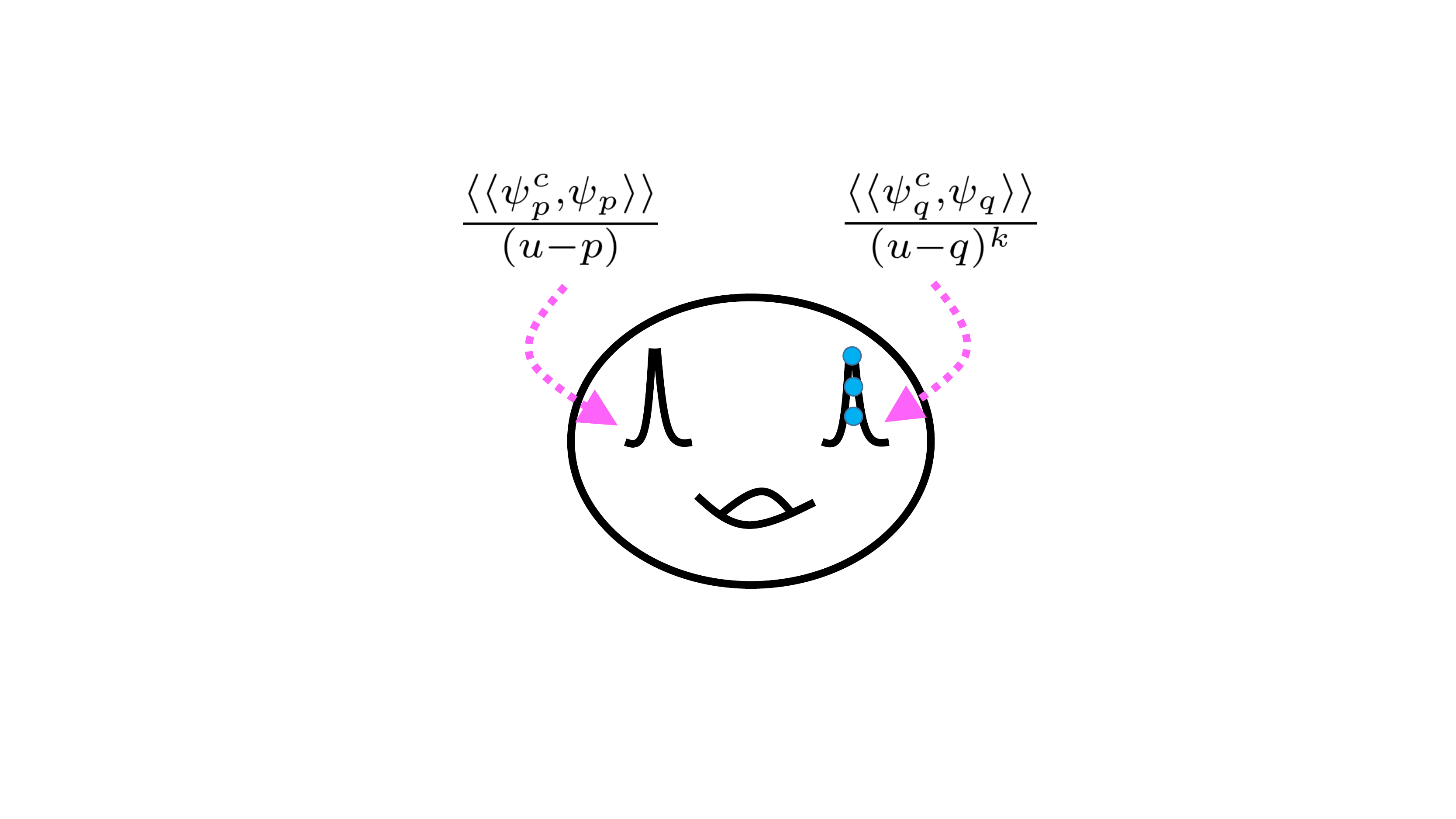}
\caption{Depiction of a Hitchin system on a genus one curve with poles at
marked points indicated by narrow cylindrical regions, that is, spikes. Background
values for localized matter fields induce poles in the Higgs field of the Hitchin system.
On the left we depict matter localized at $u = p$ which generates
a simple pole. On the right we depict matter localized at the non-reduced scheme $(u-q)^k = 0$
which generates a higher order pole.}
\label{WildStuff}%
\end{figure}

Now, in actual physical applications, we also expect to have
matter fields localized at points of the geometry. In F-theory, these matter
fields are realized from collisions of intersecting 7-branes. Background
values for these fields lead to localized sources for
the Hitchin system equations. This is reflected in singularities (including possibly higher order singularities) for the Higgs field:
\begin{equation}
\Phi \sim du \left(\frac{T_k}{u^k} + ... + \frac{T_1}{u} \right),
\end{equation}
where $T_i$ are elements of the complexified algebra, and the singularity
is concentrated at $u=0$, with $u$ a local coordinate on the curve. The case of a simple pole, namely $k = 1$
has been studied in various contexts, and is known in the Hitchin system literature as a regular singularity. This
case is particularly tractable because the data of the singularity is fully captured by a residue around $u = 0$.  Higher order poles are often
referred to as irregular or wild. This case is more delicate because a residue will fail to detect higher order terms.
T-brane configurations correspond to cases where any or all of the $T_i$ are actually nilpotent.

Physical considerations impose limits on possible T-brane phenomena.
Such constraints arise because the total number of matter fields in a
string compactification often obeys additional conditions beyond those imposed by the local Hitchin system.
In this paper we provide a physical picture from string compactification for such singularities. Moreover, we will
show the sense in which the structure of possible singularities is constrained by the additional
assumption that they arise from a genuine string compactification.

The essential point is that these singularities are all induced from background values for localized matter
fields. The case of higher order singularities involves a particular subtlety in
F-theory compactification which as far as we are aware, has not been addressed previously.
Much as in earlier work on localized matter in
F-theory (see e.g. \cite{TBRANES, BHVI, KatzVafa}), we consider a Hitchin system with gauge group $G_{\mathrm{parent}}$.
Activating a background value for the Higgs field $\Phi_{\mathrm{parent}}$ initiates a breaking pattern
to a lower rank gauge group $G$, with matter fields in various irreducible representations of $G$. Schematically,
localized matter fields are associated with elements in the ring:
\begin{equation}
\frac{\mathbb{C}[u]}{(\alpha_{R}(u))},
\end{equation}
where   $\alpha_{R}(u)$ serves to remind us that the localization depends on the choice of
representation $R$ for a given matter field. Now, in the generic case,
$\alpha_{R}$ has a simple zero, and this  corresponds to the case of a single localized matter field.
When we have higher order zeros, we obtain additional matter
fields localized at the same point. This is possible because strictly speaking, there is a difference
between the point defined by $u = 0$ and that defined by $u^k = 0$. In the latter case, we have what is sometimes referred to as a
non-reduced scheme. Additional structure lurks in such objects, which as we argue is crucial in developing a
consistent picture for how localized matter appears in F-theory compactifications. Indeed,
background values for these higher order matter fields translate to higher order poles for the Higgs field of the
theory with gauge group $G$. Varying these background values then determines a moduli space of vacua which we
match to that of a wild Hitchin system. See Figure \ref{WildStuff} for a depiction of higher order poles induced
from background values for matter fields.

In some sense, this accomplishes the main point of matching open and closed string moduli. Indeed, since expectation values
for localized matter correspond in F-theory to complex structure deformations of the Calabi-Yau, we see that the limiting
behavior of these moduli (in tandem with the rest of the intermediate Jacobian) provide a characterization which extends
to wild Hitchin systems as well.

Physical considerations impose additional constraints. In 6D vacua, anomaly cancellation
conditions tend to impose tight restrictions on the total number of matter fields. This in turn limits the possible order of poles which
can be realized in a Hitchin system derived from intersecting 7-branes. In 4D $\mathcal{N} = 2$ vacua, the related anomaly cancellation condition
is quite innocuous, but is instead replaced by the stringent condition that gravity can be consistently decoupled.

On the other hand, the geometric formulation of these emergent Hitchin-like systems provides a single framework
to unify seemingly different moduli space problems. Indeed, we can pass to wild Higgs bundle configurations
with different orders for poles simply by adjusting the gauge singlet moduli of the F-theory model.

To test these ideas further, we also present some examples of compact F-theory geometries which realize the above considerations. In particular, we focus on the case of an $SU(2)$ Hitchin system with poles. An additional feature of these global examples is that there can often be multiple higher order singularities which must all be treated simultaneously in the Hitchin system.
Tracking through the different possible singularity types for the Hitchin system and their geometric avatars
reveals a precise match.

The rest of this paper is organized as follows. In section \ref{sec:CURVE}
we review some of the previous work on T-branes in F-theory and type IIA
compactifications on an elliptically fibered Calabi-Yau threefold.
We also extend some of these results, explaining how T-branes
can be used as a nucleation point for building more general
solutions to the Hitchin system, and consequently, the class of geometries realized
by such configurations. After this, in section \ref{sec:POINT}
we turn to the case of T-branes at a simple point, namely cases where the Higgs field develops a simple pole.
In section \ref{sec:WILD} we turn to the case of Higgs fields with higher order singularities. We derive the main equations for the Higgs field in these cases, determine the structure of the moduli space, and explain the sense in which physics imposes non-trivial constraints on the order of poles in the system. We follow this with explicit compact models in section \ref{sec:GLOBAL}. We present our conclusions and future directions for research in section \ref{sec:CONC}. In the Appendices we provide additional mathematical details on the formal structure of wild Hitchin systems and the correspondence with geometry.

\section{T-Branes on a Curve \label{sec:CURVE}}

In this section we review the realization of T-branes on a curve in the Hitchin system, and in particular, how this data can be mapped to the local moduli of a Calabi-Yau threefold defined by a curve of singularities. This correspondence actually appears in two related physical
contexts. First of all, we can consider 6D supersymmetric vacua generated by F-theory on an elliptically fibered Calabi-Yau threefold. The Hitchin system on a curve $C$ comes about as the long distance approximation for 7-branes wrapped on $C$. As already mentioned, this yields a
system with $\mathcal{N} = (1,0)$ supersymmetry, i.e. eight real supercharges. Compactifying on an additional $T^2$, we obtain type IIA string theory on the same Calabi-Yau threefold, and a 4D $\mathcal{N} = 2$ supersymmetric effective field theory.

Consider, then, F-theory on an elliptically fibered Calabi-Yau threefold $X$ with base $B$. In minimal Weierstrass form, we have:
\begin{equation}
y^2 = x^3 + fx + g,
\end{equation}
where $f$ and $g$ are sections of $\mathcal{O}_{B}(-4K_{B})$ and
$\mathcal{O}_{B}(-6K_{B})$, with $K_{B}$ the canonical class of the base $B$.
For each irreducible component $C$ of the discriminant $\Delta = 4f^3 + 27g^2$
we get a 7-brane gauge theory with gauge group $G$, as dictated
by the order of vanishing for $f$ and $g$, as well as possible
monodromic identifications in the fiber. The field content
which propagates on $C$ includes an adjoint
valued $(1,0)$ form $\Phi$, and a gauge connection $A$.
When the background values of all localized matter fields are zero,
the equations of motion for this system are \cite{HitchinSelf}:
\begin{equation}
\overline{\partial}_A \Phi = 0 \text{\,\,\,and\,\,\,} F + [\Phi,\Phi^\dag] = 0,
\end{equation}
modulo unitary gauge transformations:
\begin{equation}
\Phi \mapsto g^\dag \Phi g \text{\,\,\,and\,\,\,} A \mapsto g^\dag A g + g^\dag \overline{\partial} g.
\end{equation}
We can parameterize the moduli space of solutions using gauge invariant Casimir invariants
constructed from $\Phi$. This yields the base of the Hitchin system moduli space.
For example, in the case of an $SU(N)$ gauge theory, take
$\mathrm{Tr}(\Phi^j)$ for $j = 2,...,N$. The full hyperkahler moduli space is
then filled out by also specifying the holonomies of the gauge field $A$ along one-cycles of $C$.

In the match to Calabi-Yau geometry, the base of the Hitchin system maps to the local complex structure moduli, that is,
those moduli which can deform the singularity type of a local curve of singularities.\footnote{For some discussion
of the extension of this correspondence to the case of Calabi-Yau fourfolds, see e.g. \cite{BHVI,DWIII, Hayashi:2009ge}.}
The fiber of the moduli space is, in IIA language given by the RR moduli filling
out periods in $H^3(X,\mathbb{R}) / H^3(X,\mathbb{Z})$, the intermediate Jacobian.
A T-brane configuration corresponds to the special case where $\Phi$ is nilpotent
over all of $C$. This is clearly a rather special set of conditions to satisfy.

The match between Hitchin space degrees of freedom and localized moduli has non-trivial implications for
Calabi-Yau geometry. For example, since the moduli space of the Hitchin system (with smooth Higgs field) consists of a
single connected component, we can perform a small perturbation in such a configuration to reach one in which $\Phi$ is not nilpotent. One
way to establish the existence of a single connected component is to
work in terms of the complexified connection $\mathcal{A} = A + \Phi + \Phi^\dag$
with curvature $\mathcal{F}$ so that the Hitchin system equations become \cite{HitchinSelf}:
\begin{equation}
\mathcal{F} = 0.
\end{equation}
The existence of the match with Calabi-Yau moduli, in tandem with the existence of a single connected component means
it is enough to take limiting behavior in the Calabi-Yau complex structure moduli to
produce simple T-branes of the Hitchin system. In future sections we will consider the extension of some of these results to the case of Hitchin sytems with singularities and the associated T-branes.

\subsection{Nilpotent Nucleation}\label{Sec:nucleation}

Even though such nilpotent configurations are quite special, they provide a convenient way to generate a broad class of
explicit solutions to the Hitchin system.  To illustrate, we will construct below T-branes in 6D theories which are never physically ``rigid" (i.e. forming an isolated component of moduli space). Instead, starting from a nilpotent solution, we can perturb to more general configurations.

We construct a T-brane by holding fixed a nilpotent element $\mu$ of the complexified Lie algebra $\mathfrak{g}_{%
\mathbb{C}
}$. At the level of group theory, by a theorem of Jacobson and Morozov there exists a homomorphism%
\begin{equation}
\rho_{\mu}:\mathfrak{sl}(2,\mathbb{C})\rightarrow\mathfrak{g}_{%
\mathbb{C}
}. \label{JacMor}%
\end{equation}
taking the raising operator of $\mathfrak{sl}(2,\mathbb{C})$ to $\mu$.
In general, the image of this $\mathfrak{sl}(2,\mathbb{C})$ will take values
in a maximal subalgebra $\mathfrak{h}_{%
\mathbb{C}
}$ such that its commutant $\mathfrak{c}_{%
\mathbb{C}
}$ is the \textquotedblleft unbroken\textquotedblright\ gauge symmetry. Using this,
we can construct T-branes in two steps:

\begin{itemize}
\item First, construct a solution in the nilpotent cone for the $SU(2)$
Hitchin system.

\item Second, define a map from the $SU(2)$ Hitchin system to the Hitchin
system with gauge group $G$   induced by the homomorphism $\rho_{\mu}.$
\end{itemize}

A general theorem of Hitchin \cite{HitchTeich} ensures that there is a
corresponding solution to the Hitchin system with gauge symmetry
$\mathfrak{h}_{%
\mathbb{C}
}$. So, starting from a T-brane configuration, we sweep
out a local neighborhood in the moduli space.

In \cite{DEL}, more general moduli spaces of nilpotent $SU(2)$ Higgs fields were constructed
in which the Higgs fields were allowed to vanish at isolated points.  Even more generally, the above construction of T-branes can be extended to higher rank gauge groups whose Higgs fields can take values in nonzero nilpotent orbits of smaller dimension at isolated points.  For example, an $SU(3)$ T-brane on a curve can be constructed whose Higgs field has rank 2 at the generic point of the curve, has rank 1 at isolated points, and vanishes at another set of isolated points.

\subsubsection{Heterotic Dual}\label{Sec:hetdual_symmcount}

It is also instructive to study the structure of this `nucleation' in the heterotic dual.
Recall that in 6D Heterotic / F-theory duality, stable holomorphic vector bundles on an elliptically fibered
K3 surface correspond to elliptically fibered Calabi-Yau
threefolds with base a Hirzebruch surface $\mathbb{F}_n$ with $-12 \leq n \leq 12$.\footnote{We are using the standard convention of F-theory whereby $\mathbb{F}_n$ means $\mathbb{F}_{-n}$ if $n<0$.  This convention is made to distinguish
between the situations where the section of $\mathbb{F}_n$ on which a gauge group is placed has positive or negative self-intersection.}  From this perspective, we would
like to verify that starting from a stable holomorphic vector bundle $V$ with structure group $SU(2)$, we can construct a
holomorphic vector bundle with more general structure group $G \subset E_8$. In particular, we wish to verify that there
are deformation moduli available which can connect this solution to one with generic values of complex structure in
the associated spectral cover. Phrased differently, we will use embeddings of the bundle structure group $SU(2) \subset E_8$ to probe\footnote{Note that the notion of using simple bundles as a ``probe" of more general bundle moduli spaces has also been successfully employed in 4D heterotic compactifications \cite{Anderson:2011ns,Anderson:2013qca,Anderson:2014hia}.} the general moduli space of $G$-bundles over $K3$.

Assuming a particular embedding of $SU(2) \rightarrow E_8$, the adjoint representation of $E_8$ will
decompose into various symmetric powers of the fundamental representation of $SU(2)$. These additional representations
specify smoothing deformations which take us from the original embedding of the vector bundle to a more general
vector bundle with structure group $G$. It should be noted that the exact structure group obtainable will be determined by $c_2(V)$ and a complete description of those $G$-bundle moduli spaces here is beyond the scope of the present work. It would be interesting to fully classify this structure in future work (see e.g. \cite{Bershadsky:1996nh} for more details on the possible moduli spaces). We now establish that \emph{some} such smoothing deformations always exist. Said differently,
our aim is to count the number of zero modes associated with
various symmetric powers of $S^{j}V$.

Since we are assuming $V$ is a stable
vector bundle with a non-trivial instanton number, we have that:%
\begin{equation}
\int_{K3}c_{2}(V)=12+n,
\end{equation}
where $-8\leq n\leq12.$ The reason for the lower bound $-8$ is that we are
assuming we can performing a breaking pattern down to $E_{7}$. For supersymmetric
vacua there is also an upper bound of at most $24$ instantons.
As $V$ is stable, we also have:%
\begin{equation}
h^{0}(K3,S^{j}V)=h^{2}(K3,S^{j}V)=0\text{ \ \ for \ \ }j>0\text{.}%
\end{equation}
As a consequence, the index theorem counts all the zero modes coming from
$S^{j}V$:%
\begin{equation}
-h^{1}(K3,S^{j}V)=\int_{K3}\text{ch}(S^{j}V)\text{Td}(K3)=\text{rk}%
(S^{j}V)\chi(K3,\mathcal{O}_{K3})+\int_{K3}\text{ch}_{2}(S^{j}V)
\end{equation}
or:%
\begin{equation}
-h^{1}(K3,S^{j}V)=2(j+1)+\int_{K3}\text{ch}_{2}(S^{j}V).
\end{equation}
Our task therefore reduces to calculating the second Chern character of
$S^{j}V$. Here, we use the splitting principle. For some line bundle $L$ on
$K3$, we have, for $k>0$ an integer:%
\begin{align}
\text{ch}_{2}(V)  &  =\text{ch}_{2}(L\oplus L^{-1})\\
\text{ch}_{2}(S^{2k}V)  &  =\text{ch}_{2}(L^{2k}\oplus L^{2k-2}...\oplus
\mathcal{O}_{K3} \oplus ...\oplus L^{2-2k}\oplus L^{-2k})\\
&  =\text{ch}_{2}(L^{2k})+\text{ch}_{2}(L^{2k-2})+...+\text{ch}_{2}%
(L^{2-2k})+\text{ch}_{2}(L^{-2k})\\
\text{ch}_{2}(S^{2k+1}V)  &  =\text{ch}_{2}(L^{2k+1}\oplus L^{2k-1}...\oplus L\oplus
L^{-1}\oplus...\oplus L^{1-2k}\oplus L^{-2k-1})\\
&  =\text{ch}_{2}(L^{2k+1})+\text{ch}_{2}(L^{2k-1})+...+\text{ch}_{2}(L^{1-2k})+\text{ch}_{2}(L^{-2k-1}).
\end{align}
In other words, we get:%
\begin{equation}
\text{ch}_{2}(S^{2k}V)=\underset{m=1}{\overset{k}{\sum}}4m^{2}c_{1}(L)^{2}%
=\frac{2(2k+1)(k+1)k}{3}c_{1}(L)^{2}.
\end{equation}
Returning to our computation of the dimension   $h^{1}(K3,S^{j}V)$ therefore yields:
\begin{equation}
-h^{1}(K3,S^{2k}V)=2(2k+1)+\frac{2(2k+1)(k+1)k}{3}\int_{K3}c_{1}(L)^{2}.
\end{equation}
On the other hand, we also know that the instanton number of the vector bundle
is set by:%
\begin{equation}
\int_{K3}c_{1}(L)^{2}=\int_{K3}\text{ch}_{2}(V)=-\int_{K3}c_{2}(V)=-(12+n).
\end{equation}
Hence, we get:%
\begin{equation}
h^{1}(K3,S^{2k}V)=\left(\frac{2k+1}3\right)\left(2k(k+1)(12+n)-6\right).
\end{equation}
Similarly, for odd symmetric powers we get
\begin{equation}
\text{ch}_{2}(S^{2k+1}V)=\underset{m=0}{\overset{k}{\sum}}(2m+1)^{2}c_{1}(L)^{2}%
=\frac{(2k+3)(2k+1)(k+1)}{3}c_{1}(L)^{2},
\end{equation}
which leads to
\begin{equation}
h^{1}(K3,S^{2k+1}V)=\left(\frac{k+1}{3}\right)\left((2k+3)(2k+1)(12+n)-12\right)
\end{equation}
as above.

The important point for us is that the deformation moduli of our vacuum
configuration have $j>1$, and thus in particular we always have $S^2 V = End_0(V)$. As $j=2$ sets a
lower bound for the dimension   $h^{1}(K3,S^{j}V)$, we have:%
\begin{equation}
h^{1}(K3,S^{j}V)\geq42+4n>0,
\end{equation}
where in the rightmost inequality we used the fact that $n \geq -8$. So,
we always have deformation moduli available to move us back to a non-singular configuration (in the language of the F-theory dual geometry). Note that this result implies that for the case of F-theory duals of such heterotic models, the moduli space of the induced \emph{singular} Hitchin systems can also be connected (see \cite{Aspinwall:1998he,Anderson:2013rka} for examples of such dual heterotic/F-theory pairs).

\section{T-Branes at a Simple Point \label{sec:POINT}}

In the previous section we focused on the case of T-brane phenomena for a genus $g$ Riemann surface. Now, in physical
realizations, it is also quite common that the fields of the Hitchin system may develop singularities at points of this
Riemann surface. To illustrate, observe that without any such singularities, $\Phi$ is an adjoint valued $(1,0)$ form,
so the Casimir invariants $\mathrm{Tr}(\Phi^n)$ will be holomorphic sections of the bundle $K_C^n$,
i.e. elements in $H^0(C , K_C^n)$. On the other hand, it is also common in physical applications for the Riemann surface to be
a $\mathbb{P}^1$ for which $H^0(C , K_C^n) = 0$. In these cases,
the Hitchin system becomes non-trivial via the fact that a $(1,0)$-form
or a higher differential on $\mathbb{P}^1$ can develop poles at various points of the curve.
This is in fact a general occurrence in F-theory.
Examples include elliptically fibered Calabi-Yau threefolds with base a Hirzebruch surface.

Now, another closely related feature of physical models is the presence of matter localized at points of the geometry. In the context of
6D theories, these matter fields fill out 6D hypermultiplets which transform in some representation $R$ of the gauge group $G$. When the
representation is pseudo-real, it is also possible to have half hypermultiplets. For a hypermultiplet, we have a pair of scalars $\psi \oplus \psi^c$, where the first scalar transforms in the representation $R$ and the second transforms in the conjugate (i.e., dual) representation $R^c$. In the associated Calabi-Yau geometry, localized matter fields are often interpreted near the
collision of distinct components of the discriminant locus.\footnote{Of course, this presupposes that geometry is an accurate guide to the matter spectrum, a point which can be obscured by T-branes \cite{TBRANES}!}

There is a close interplay between the background values for these hypermultiplet scalars and possible polar terms in the Higgs field. Indeed,
the holomorphic F-term data of the Hitchin system now receives the correction term (see e.g. \cite{BHVI}):
\begin{equation}\label{SimplePole}
\overline{\partial}_A \Phi = \delta_{p} \langle \langle \psi^c , \psi \rangle \rangle
\end{equation}
where $\delta_{p}$ is a delta function (namely, a $(1,1)$ current) localized at the point $u = p$. Here, we have also introduced the canonical pairing with image in the adjoint representation of the complexified algebra:
\begin{equation}
\langle \langle \cdot , \cdot \rangle \rangle : R^c \otimes R \rightarrow \mathrm{ad}(\mathfrak{g}_\mathbb{C}).
\end{equation}
In the context of colliding 7-branes, it can happen that there are actually multiple hypermultiplets all concentrated at the same point. For example, in the collision of an $SU(N)$ 7-brane with an $SU(M)$ 7-brane, the hypermultiplets transform in the bifundamental representation $(\mathbf{N},\overline{\mathbf{M}})$, so from the perspective of the $SU(N)$ gauge theory we actually have $M$ hypermultiplets in the fundamental representation of $SU(N)$. Let us also note that it is not even necessary to have weakly coupled matter fields. Strongly coupled generalizations of such hypermultiplets known as conformal matter generate the same
sort of deformations of the Hitchin system \cite{DelZotto:2014hpa, Heckman:2014qba, Heckman:2016ssk, Mekareeya:2016yal}.
In this more general formulation, we simply have a source term
sitting on the right hand side of equation (\ref{SimplePole}).
Assuming such a source term is present, and denoting by ``...'' the regular terms, integrating equation (\ref{SimplePole}) yields:
\begin{equation}
\Phi \sim du \frac{\langle \langle \psi^c , \psi \rangle \rangle}{u - p} + ...
\end{equation}

In physical constructions, one typically has multiple marked points,
each with localized matter. When this matter has a non-zero background
value, we obtain a parabolic Higgs bundle. See Appendix \ref{app:wild} for review
of some aspects of this case.\footnote{For recent work on parahoric Hitchin systems and the corresponding integrable systems,
see \cite{PARAHORIC}.} Holding fixed a choice of boundary conditions
at each such marked point, we can then construct a corresponding moduli
space for the Hitchin system. Here, the gauge invariant data of the boundary
condition is captured by the conjugacy class in $\mathfrak{g}_\mathbb{C}$
of the residue. Of course, in the full physical construction
we are free to vary the background values of the hypermultiplets $\psi \oplus \psi^c$,
and in so doing change the boundary conditions for the parabolic Higgs bundle.

For a given choice of background fields at a marked point, we obtain a nilpotent
element $\mu \in \mathfrak{g}_{\mathbb{C}}$. The conjugacy class is then specified by the nilpotent
orbit of this element. This non-zero background value also initiates a breaking pattern of $\mathfrak{g}_{\mathbb{C}}$
to a commutant subalgebra which we denote by $\mathfrak{c}_{\mathbb{C}}$. Roughly speaking, the more hypermultiplets
with non-zero background values, the lower the rank of $\mathfrak{c}_{\mathbb{C}}$.
The precise breaking pattern of course depends on the
specific representations in question, and is best addressed using
the Bala-Carter theory of nilpotent orbits (see e.g. \cite{NilpotentBook}).

This data is hidden from the complex structure of the local Calabi-Yau geometry. Just as in reference \cite{Anderson:2013rka},
we can start from a nilpotent element, and the corresponding raising operator $T_+$ of the associated $\mathfrak{su}(2)$ subalgebra.
Perturbing by the lowering operator $T_- = T_+^{\dag}$,\footnote{Strictly speaking, this perturbation as we have described it may only make sense locally along the curve.} we obtain a family of diagonalizable deformations:
\begin{equation}\label{TsmoothBsmoove}
T(\varepsilon)=T_{+}+\varepsilon T_{-}.
\end{equation}
An interesting feature of this procedure is that the closure of the conjugacy class can
indeed jump between the cases $\varepsilon = 0$ and $\varepsilon \neq 0$. Let us note that
examples of this type include minimal rigid nilpotent orbits. Such boundary conditions are important
in the context of geometric Langlands duality and rigid surface operators \cite{Gukov:2006jk}.

Indeed, if we are only interested in the parabolic Hitchin system, each
choice of conjugacy class labels a distinct component of the moduli space.
Physically, however, we recognize that these different choices of boundary conditions are connected
to one another by activating background values of localized matter \cite{BHVI}.
One can view the results of the present paper as a general method for geometrically engineering
various surface operators, but in which we extend the moduli space by promoting some
boundary conditions to dynamical fields.

Our plan in the rest of this section will be to present some examples of T-branes at a simple point.
In particular, we shall focus on the case of minimal nilpotent orbits, namely those cases where the commutant
subalgebra has maximal rank. For all simple algebras other than $\mathfrak{e}_8$, this is realized via a non-zero background
value for a single hypermultiplet in the fundamental representation of the algebra.
In the case of $\mathfrak{e}_8$, the analogue of localized matter fields is instead played by conformal
matter fields, namely, the Higgs branch of heterotic small instantons. We revisit this example
in subsection \ref{ssec:TAN}.

\subsection{Minimal Nilpotent Orbits: Classical Algebras}

To illustrate the general idea, we begin by constructing the minimal nilpotent orbits when the gauge group
$G$ is a simple classical algebra, namely the cases of the $SU(N)$, $Sp(2N)$ and $SO(2N)$ algebras. Since the latter two cases arise
in string constructions from adding orientifolds and/or monodromic quotients to the $SU(N)$ case, we shall primarily confine our discussion
to the geometric realization of minimal nilpotent $SU(N)$ algebras.

Recall that we are interested in constructing a T-brane configuration such that the commutant subalgebra has maximal rank. The relevant
decomposition into subalgebras is:
\begin{align}
\mathfrak{su}(N)  &  \supset \mathfrak{su}(2) \times \mathfrak{su}(N-2)\times \mathfrak{u}(1)\\
\mathfrak{sp}(2N)  &  \supset \mathfrak{su}(2) \times \mathfrak{sp}(2N-2)\\
\mathfrak{so}(2N)  &  \supset \mathfrak{su}(2) \times \mathfrak{so}(2N-4) \times \mathfrak{u}(1)\\
\mathfrak{so}(2N + 1)  &  \supset \mathfrak{su}(2) \times \mathfrak{so}(2N-3) \times \mathfrak{u}(1),
\end{align}
where the T-brane is embedded in the $\mathfrak{su}(2)$ factor.
Referring back to equation (\ref{TsmoothBsmoove}), let
us note that for all cases other than the $\mathfrak{su}(N)$ example, the
closure of the conjugacy classes for $\varepsilon = 0$ and $\varepsilon \neq 0$ are different.

We shall now turn to the geometric realization of these deformations, at least for the case $\varepsilon \neq 0$.
For ease of exposition, we focus on the case of the $\mathfrak{su}$ algebra. Similar considerations hold for the other cases,
using for example the spectral curve of the associated Hitchin system. The local presentation of the Calabi-Yau threefold
is given by a curve of A-type singularities. We can write this as:
\begin{equation}
y^2 = x^2 + u^N + \varepsilon u^{N - 2}.
\end{equation}
We realize a T-brane by taking a limit with $\varepsilon \rightarrow 0 $. The analysis of this case is rather similar to what is
presented in reference \cite{Anderson:2013rka}.

\subsection{Minimal Nilpotent Orbits: Exceptional Algebras}

In  the case of the exceptional algebras,  the relevant
decomposition into subalgebras is:
\begin{align}
\mathfrak{e}_{8} &  \supset\mathfrak{e}_{7}\times\mathfrak{su}(2)\\
\mathfrak{e}_{7} &  \supset\mathfrak{so}(12)\times\mathfrak{su}(2)\\
\mathfrak{e}_{6} &  \supset\mathfrak{su}(6)\times\mathfrak{su}(2)\\
\mathfrak{f}_{4} &  \supset\mathfrak{sp}(6)\times\mathfrak{su}(2)\\
\mathfrak{g}_{2} &  \supset\mathfrak{su}(2)\times\mathfrak{su}(2),
\end{align}
where the T-brane is embedded in the $\mathfrak{su}(2)$ subalgebra.
The F-theory realization of these $\varepsilon$-deformed T-brane
configurations is:
\begin{align}
\mathfrak{e}_{8}  &  :y^{2}=x^{3}+u^{5}+\varepsilon xu^{3}\\
\mathfrak{e}_{7}  &  :y^{2}=x^{3}+xu^{3}+\varepsilon x^{2}u\\
\mathfrak{e}_{6}  &  :y^{2}=x^{3}+u^{4}+\varepsilon xu^{2}\\
\mathfrak{f}_{4}  &  :y^{2}=x^{3}+q u^{4}+\varepsilon xu^{2} \\
\mathfrak{g}_{2}  &  :y^{2}=x^{3}+q xu^{2}+\varepsilon u^{2}  .
\end{align}
The factors of $q$ in the non-simply laced cases are introduced in order to
pass to the non-split type of each elliptic fiber \cite{BershadskyPLUS}.

\section{T-Branes Gone Wild \label{sec:WILD}}

In this section we consider a more general class of T-brane configurations which originate from allowing
$\Phi$ to develop higher order poles. More precisely, we now ask whether we can realize
a Higgs field of the form:
\begin{equation}\label{PhiWild}
\Phi = du \left(\frac{T_k}{u^k} + ... + \frac{T_1}{u}  + \ldots\right)
\end{equation}
where the rightmost set of ``...'' refers to regular terms in the Higgs field. Here, the generators $T_k$ take values
in $\mathfrak{g}_\mathbb{C}$, the complexification of the gauge algebra for the Hitchin system.

Our plan will be to geometrically engineer such configurations via colliding 7-branes. In particular,
we will argue that just as in the case of simple poles, these higher order residues can be understood as background values of
matter fields. The main distinction compared with the case of simple poles is that now we allow
matter to be localized at a non-reduced scheme $u^k = 0$. Another goal will be to understand
how constraints from anomaly cancellation (in the case of 6D F-theory vacua) or the
condition that a global model exists (in the case of 4D type IIA vacua) leads to a non-trivial upper bound on the
singular behavior possible in such configurations.

The moduli space of solutions for the Hitchin system with wild ramification (i.e. an irregularity singularity)
is quite subtle, and is the subject of much   work in the mathematics and physical
mathematics literature, and originated with the work of Boalch
(for a general survey see e.g. \cite{BoalchHabil} and
references therein). In Appendix \ref{app:wild} we present a brief overview
of some of these results in the case of $SU(2)$ gauge theory with poles
of order up to four.

To briefly illustrate some of these subtleties, consider the parameterization of the Hitchin system
in terms of the complexified connection $\mathcal{A} = A + \Phi + \Phi^\dag$. The complexified
connection will also have poles at the same locations as $\Phi$. The first issue is that although the
holonomy of $\mathcal{A}$ detects first order poles (via a residue theorem), higher order poles are not purely topological
in form, but appear to depend on a choice of coordinate system near the marked point. Indeed, observe that a
complexified gauge transformation:
\begin{equation}
(d + \mathcal{A}) \mapsto g_{\mathbb{C}}^{-1} \cdot (d + \mathcal{A}) \cdot g_{\mathbb{C}},
\end{equation}
can shift the order of higher order poles provided we allow $g_{\mathbb{C}}$ to also be singular at $u$.
To deal with such issues, we need to have a more precise notion of which types of singular behavior one should allow.

In  physical applications, we can fix some of these ambiguities by requiring that all localized matter fields
in the associated geometry remain normalizable. To illustrate, consider a 4D F-theory vacuum containing a
7-brane gauge theory wrapped on the K\"ahler surface $C \times T^2$. In this system, we can have matter fields
localized on either the factor $C$ or the factor $T^2$. Consider, then, a matter field which is localized at a point of $T^2$,
but which transforms as a holomorphic section of a bundle defined on $C$. Following the discussion presented in
\cite{Witten:2007td}, such matter fields obey an equation of the schematic form:
\begin{equation}\label{holosolve}
\left(\frac{\partial}{\partial u} + \mathcal{A}_u \right) \cdot \Psi = 0,
\end{equation}
where we assume $\Psi$ transforms in a representation $R$ of the gauge group.
The presence of the singularity at $u = 0$ means that we must exercise
care in writing the normalizable solutions to this equation. For example, we can
formally solve equation (\ref{holosolve}) to find:
\begin{equation}
\Psi^{(i)} \sim \exp\left(\frac{a^{(i)}_k}{u^{k-1}} + ... \right)\label{holaPhi}
\end{equation}
for some $a^{(i)}_k$. Here, the superscript $(i)$ labels one component
in the vector defined by $\Psi$ in the representation $R$.

The solution  \eqref{holaPhi} is only normalizable in the sector of the (complex) $u$-plane
where $\mathrm{Re}({a^{(i)}_k}/{u^{k-1}}) < 0$. When we pass
to another sector, we must take a linear combination of the solutions
in this sector to obtain another solution. Following Boalch's work, there are precisely $2(k-1)$
such sectors, i.e. Stokes chambers, and for each one
we get a transition matrix from chamber $i$ to chamber $i+1$,
which we denote by $S_i$. The moduli space problem of interest
will then involve holding fixed the generalized monodromy:
\begin{equation}
\widehat{M} = \exp(2 \pi i T_1) \cdot S_{1} S_{2} ... S_{2(k-1)}.
\end{equation}
We refer to deformations which hold fixed this data as isomonodromic.
Note that the $T_k$ of equation (\ref{PhiWild}) are still free to vary.
For additional details on the theory of isomonodromic deformations
of meromorphic differential equations, see for example \cite{JIMBO}.

Our plan in this section will be to show how to engineer wild T-branes in F-theory. Our main result is that
these higher order poles are generated by matter fields localized at non-reduced schemes such as $u^k = 0$.
In this sense, it simply requires additional tuning in the complex structure moduli of an F-theory compactification to
realize these more subtle configurations. Now, precisely because this deformation problem is captured by the moduli space
of a local Calabi-Yau, constraints from anomaly cancellation bound the number of
such matter fields. Moreover, further constraints arise if we attempt to embed the local model in a globally complete geometry.
All told, this greatly limits the possible configurations of wild T-branes, including
the total order of poles, as well as the possible values of the generalized residues $T_i$ which can actually be engineered.
To illustrate, we calculate both the physical moduli space (as defined by F-theory) as well as that defined by
the wild Hitchin system of isomonodromic deformations.

\subsection{Wild Matter}

We shall now turn to the way in which higher order poles in the Higgs field can arise.
To this end, let us consider in more detail the way in which we generate localized modes from the perspective of
an 8D 7-brane gauge theory. Along these lines, it is again helpful
to work in terms of a 7-brane wrapping a K\"ahler surface $S= C \times T^{2}$,
i.e., we compactify our 6D theory on an additional $T^{2}$ to
four dimensions. Our plan will be to study localized matter fields obtained by Higgsing a parent gauge theory
defined on a patch of this K\"ahler surface. We use a local coordinate $u$ for $C$ and $v$ for the $T^2$ factor.
Much as in earlier work on modelling intersecting 7-branes using this 8D gauge theory, matter fields will arise from localized
vortex equations. The key difference from earlier work will be in the profile for the parent gauge theory Higgs field we use
to trap matter along a non-reduced scheme.

Considering a 7-brane wrapped on a K\"ahler surface, we can parameterize the
higher Kaluza-Klein modes of the system in terms of a collection of 4D $\mathcal{N} = 1$
superfields \cite{BHVI} (see also \cite{SiegelTEND, WackerGregoire, DijkgraafVafaIII}).
The resulting supersymmetric equations of motion for the system dictate
the profiles of the internal fields. To capture the main features of higher order poles in the Higgs
field, it is enough to track the F-term equations of motion, modulo complexified gauge transformations. That is,
we shall exclusively work in holomorphic gauge. This will make the match with complex
geometry especially transparent, and with no loss of generality.\footnote{The
passage back to a unitary frame where we impose F- and D-terms
modulo unitary gauge transformations is achieved by a suitable
complexified gauge transformation (see e.g. \cite{TBRANES, FGUTSNC}).}
For a 7-brane with no localized matter, the F-term equations of motion are governed by the
superpotential \cite{TBRANES, BHVI, DWI, FGUTSNC}:
\begin{equation}
W_{\text{bulk}}=\underset{S}{\int}\text{Tr}(\Phi_{(2,0)}\wedge F_{(0,2)}).
\label{bulksuper}%
\end{equation}
The first order equations of motion for this system are:%
\begin{equation}
\overline{\partial}_{A}\Phi=0\text{ \ \ and \ \ }F_{(0,2)}=0\text{.}%
\end{equation}
Next, expand around a specific background $A^{(0)}$ and $\Phi^{(0)}$ which
satisfies these equations of motion,  allowing $\Phi^{(0)}$ to possibly have poles along a divisor of $S$. Expanding around this
background, we write:
\begin{align}
A  &  =A^{(0)}+A^{(1)},\\
\Phi &  =\Phi^{(0)}+\Phi^{(1)}.
\end{align}
Plugging this into the original system of equations, we obtain the
first order F-term relations:%
\begin{equation}
\overline{\partial}_{A^{(0)}}\Phi^{(1)}+[A^{(1)},\Phi^{(0)}]=0\text{ \ \ and
\ \ }\overline{\partial}_{A^{(0)}}A^{(1)}=0.
\end{equation}
Following \cite{FGUTSNC}, since $\left(  \overline{\partial}_{A^{(0)}%
}\right)  ^{2}=0$ we can express our
solution in a local gauge as:%
\begin{equation}
\Phi^{(1)}=[\xi,\Phi^{(0)}]+h\text{ \ \ and \ \ }A^{(1)}=\overline{\partial
}_{A^{(0)}}\xi\text{,}%
\end{equation}
for $h$  a holomorphic $(2,0)$ form valued in ad$P$
with $P$ a principal $G_{\text{parent}}$-bundle, and $\xi$ a
$(0,0)$ form valued in ad$P$.

To proceed further, we now assume a specific form for $\Phi^{(0)}$. In terms of the local
coordinates $u$ and $v$ introduced previously, $\Phi^{(0)}$ takes the form:
\begin{equation}
\Phi^{(0)}=\phi \, du\wedge dv\text{,}%
\end{equation}
for $\phi$ an adjoint valued scalar in the complexification of
$\mathfrak{g}_{\mathbb{C}}^{\text{parent}}$. Denote the adjoint action by
$\phi$ as $\mathrm{ad}_{\phi}$. In this case, we can make a further decomposition of
the adjoint action according to the decomposition into irreducible
representations of the unbroken gauge group. Assuming that we have a parent
gauge group $G_{\text{parent}}$ which breaks to $G$ (which may contain
multiple semi-simple factors), we have a further decomposition into irreducible
representations of the original adjoint representation:%
\begin{equation}
\text{ad}(G_{\text{parent}})=\underset{i}{\oplus}R_{i}\text{.}%
\end{equation}
For each such irreducible representation, denote the
eigenvalue of $\mathrm{ad}_{\phi}$ by $\alpha_{R}$. The resulting
matter fields transforming in a representation $R$ of $G$ then satisfy the
equations:%
\begin{equation} \label{Pantera}
\Phi_{R}^{(1)}=\alpha_{R}\xi_{R}+h_{R}\text{ \ \ and \ \ }A_{R}^{(1)}%
=\overline{\partial}_{A^{(0)}}\xi_{R}\text{,}%
\end{equation}
in the obvious notation. We can then present localized solutions as:%
\begin{equation}
\Phi_{R}^{(1)}=\alpha_{R}\xi_{R}+h_{R}\text{ \ \ and \ \ }A_{R}^{(1)}=\overline
{\partial}_{A^{(0)}}\left(  \frac{\Phi_{R}^{(1)}-h_{R}}{\alpha_{R}}\right)
\text{.}%
\end{equation}

We obtain a class of solutions by taking $\phi$ valued in the
Cartan subalgebra with simple zeros. For example, we can consider the
breaking pattern induced by taking:%
\begin{equation}
\phi=\left[
\begin{array}
[c]{cc}%
Mu1_{N\times N} & \\
& -Nu1_{M\times M}%
\end{array}
\right]  \text{.}%
\end{equation}
In this case, we have localized modes in the bifundamental representation of
$SU(N)\times SU(M)$, which are trapped at $u=0$:%
\begin{equation}
\Phi_{N\times M}^{(1)}=(M+N)u\xi_{N\times M}+h_{N\times M}\text{ \ \ and
\ \ }A_{N\times M}^{(1)}=\overline{\partial}_{A^{(0)}}\left(  \frac
{\Phi_{N\times M}^{(1)}-h_{N\times M}}{u}\right)  \text{,}%
\end{equation}
where  the subscript $R$ denotes the representation with respect to the
gauge group left unbroken by the background choice of $\phi$.

We can also entertain more general polynomials in $u$:%
\begin{equation}
\phi=\left[
\begin{array}
[c]{cc}%
M\alpha_{R}(u)1_{N\times N} & \\
& -N\alpha_{R}(u)1_{M\times M}%
\end{array}
\right]  \text{,}%
\end{equation}
which yields the zero modes:%
\begin{equation}
\Phi_{N\times M}^{(1)}=(M+N)\alpha_{R}(u)\xi_{N\times M}+h_{N\times M}\text{
\ \ and \ \ }A_{N\times M}^{(1)}=\overline{\partial}_{A^{(0)}}\left(
\frac{\Phi_{N\times M}^{(1)}-h_{N\times M}}{\alpha_{R}(u)}\right)  \text{.}%
\end{equation}
Provided $\alpha_{R}(u)$ has simple zeros, we get localized matter in the
bifundamental of $SU(N)\times SU(M)$. If, however, multiple zeroes coincide,
we instead obtain a higher order pole.

Returning to the general thread of our discussion, we see that the localized modes
which descend from bulk modes are captured in terms of the quantity $\xi$, which has the local
expression:
\begin{equation}\label{xiexpress}
\xi_{R} = \frac{\psi(u)}{\alpha_R(u)}, \,\,\,\text{and}\,\,\,\xi_{R^c} = \frac{\psi^c(u)}{\alpha_{R}(u)}.
\end{equation}
Here, we have used the fact that a full
hypermultiplet localizes together (see Appendix B of
reference \cite{TBRANES}). The general statement,
then, is that for hypermultiplet matter $\psi\oplus
\psi^{c}$ localized at the zeroes of $\alpha_{R}(u)$, we have local
representatives in:
\begin{equation}
\psi\in K_{T^2}^{1/2} \otimes R\otimes\frac{\mathbb{C}[u]}{\left(  \alpha_{R}(u)\right)  }\text{
\ \ and \ \ }\psi^{c}\in K_{T^2}^{1/2} \otimes R^{c}\otimes\frac{\mathbb{C}[u]}{\left(  \alpha
_{R}(u)\right)  },
\label{localreps}
\end{equation}
that is, we can write down power series expansions:%
\begin{equation}
\psi(u)=\underset{i=0}{\overset{k-1}{%
{\displaystyle\sum}
}}\psi_{i}u^{i}\text{ \ \ and \ \ }\psi^{c}(u)=\underset{i=0}{\overset{k-1}{%
{\displaystyle\sum}
}}\psi_{i}^{c}u^{i}.
\end{equation}
In the above expressions, we note in passing that both $\psi$ and $\psi^c$ also transform as spinors on the
matter curve $T^2$ factor, i.e. we have included a factor of $K_{T^2}^{1/2}$.

Consider now the coupling of these localized modes to the other bulk degrees
of freedom of the system. For a local model with matter generated by
$\alpha_{R}=u^{k}$, we get $k$ zero modes all localized at $u=0$. Therefore, plugging into
our bulk superpotential, we can read off the coupling of the bulk gauge field
to these boundary modes:
\begin{equation}\label{WTtwo}
W_{T^{2}}=\underset{S}{\int}\overline{\partial}_{\overline{u}}\left(
\Phi^{(1)}_{R^c}\cdot(\overline{\partial}_{\overline{v}}+A_{\overline{v}})\cdot A^{(1)}_{R}
- \Phi^{(1)}_{R}\cdot(\overline{\partial}_{\overline{v}}+A_{\overline{v}})\cdot A^{(1)}_{R^c} \right),
\end{equation}
where $u$ is a local coordinate transverse to the matter curve and $v$
is a local coordinate along the matter curve. In this expression, we have also
kept implicit the pairing with respect to just one of the simple gauge group
factors, namely the one localized on $S= C \times T^{2}$. We trace over the
representation content of the other gauge group factors. As an example of this
procedure, consider the case of $G=SU(N)$, with each matter field a
bifundamental of $SU(N)\times SU(M)$. In this case, we trace over the flavor
index, i.e. the index of $SU(M)$, whilst allowing a non-trivial covariant
derivative of $SU(N)$ to act on the localized matter fields.

In arriving at (\ref{WTtwo}) we have used the fact that
there is a natural symplectic pairing between the scalars of the
hypermultiplet \cite{TBRANES}. Inserting our expressions for the localized
fluctuations $\Phi_{R}^{(1)}$ and $A_{R}^{(1)}$ from (\ref{Pantera}) and (\ref{xiexpress}) into $W_{T^2}$ then yields:
\begin{equation}\label{WTtwoagain}
W_{T^{2}}=\underset{S}{\int}\overline{\partial}_{\overline{u}}\left(
\frac{\psi^{c}(u)\cdot(\overline{\partial}_{\overline{v}}+A_{\overline{v}%
})\cdot\psi(u)}{u^{k}}\right).
\end{equation}
Note that in the above, we have a power series expansion in $u$ for
$\psi(u)$ and $\psi^{c}(u)$. So, although the product of $\psi(u)$ and
$\psi^{c}(u)$ has terms of degree zero to degree $2(k-1)$, the only terms
which actually survive are from degree zero to degree $k-1$, the higher order
terms being regular, and thus annihilated by $\overline{\partial}%
_{\overline{u}}$. Another feature of this formula is that there is a non-degenerate pairing of the simple poles. Such
terms can be evaluated via a residue integral, and yield standard kinetic terms on the $T^2$ factor \cite{TBRANES, FGUTSNC}.

The higher order poles present in (\ref{WTtwoagain}) are what generate
wild behavior in our Hitchin system. To see this, consider the equations
of motion obtained by varying with respect to the bulk fields of the system.
The bulk F-term equations of motion are now:
\begin{equation}
\overline{\partial}_{A}\Phi=\overline{\partial}_{\overline{u}}\left(
\frac{\left\langle \left\langle \psi^{c}(u),\psi(u)\right\rangle \right\rangle
}{u^{k}}\right)  \text{ \ \ and \ \ }F_{(0,2)}=0.
\end{equation}
A similar expression holds for the D-term equations of motion.
Much as in \cite{BHVI}, we have introduced a canonical pairing $\left\langle
\left\langle \psi^{c}(u),\psi(u)\right\rangle \right\rangle $ valued in $K_{T^2} \otimes \mathrm{ad} P$, with
$P$ a principal $G$ bundle. Here, we also trace over the flavor indices.
For example, in the special case of $G=SU(N)$ and a flavor group $SU(M)$, we
write, for $\alpha$ an index for the adjoint representation of $G$:%
\begin{equation}
\left\langle \left\langle \psi^{c}(u),\psi(u)\right\rangle \right\rangle
_{\alpha}=\underset{m}{\sum}\psi_{m}^{c}(u)\cdot V_{\alpha}^{(R)}\cdot\psi
_{m}(u),
\end{equation}
where $V_{\alpha}^{(R)}$ is a generator of the algebra in the
representation $R$.

Let us now collect the terms of the outer product $\left\langle \left\langle
\psi(u),\psi^{c}(u)\right\rangle \right\rangle $ in terms of a collection of
$k$ rank $M$ matrices, $T_{1},...,T_{k}$. Strictly speaking, we view the $T_j$ as
holomorphic sections of $K_{T^2}$ valued in the adjoint representation of $G$.
Since, however, $K_{T^2}$ is trivial, we can freely switch between these conventions.
Explicitly, the terms of the outer product are obtained by
expanding our power series and keeping all terms which are not regular in our
meromorphic expansion. Doing so, we arrive at the final form of the F-term
equations of motion:%
\begin{equation}
\overline{\partial}_{A}\Phi=\overline{\partial}_{\overline{u}}\left(
\frac{T_{k}+...+T_{1}u^{k-1}}{u^{k}}\right)  \text{ \ \ and \ \ }F_{(0,2)}=0.
\label{feom}
\end{equation}
One can also express the right hand side as:
\begin{equation}
\overline{\partial}_{A}\Phi= 2 \pi i \left(\delta_{u}T_{1} + \underset{j=1}{\overset{k-1}{%
{\displaystyle\sum}
}}\frac{(-1)^j}{j!}\partial_{u}^{j}\delta_{u}T_{j+1} \right).
\end{equation}
Locally, then, we have:%
\begin{equation}
\Phi\sim du \left(\frac{T_{k}}{u^{k}}+...+\frac{T_{1}}{u} + \text{regular terms at~} (u=0)\right),
\end{equation}
that is, we obtain the expected behavior of the Higgs field. Of course, we are
now free to restrict to the case of the Hitchin system, i.e., by
decompactifying the $T^{2}$ factor.

\begin{figure}[t!]%
\centering
\includegraphics[
scale = 0.50, trim = 0mm 0mm 0mm 30mm]%
{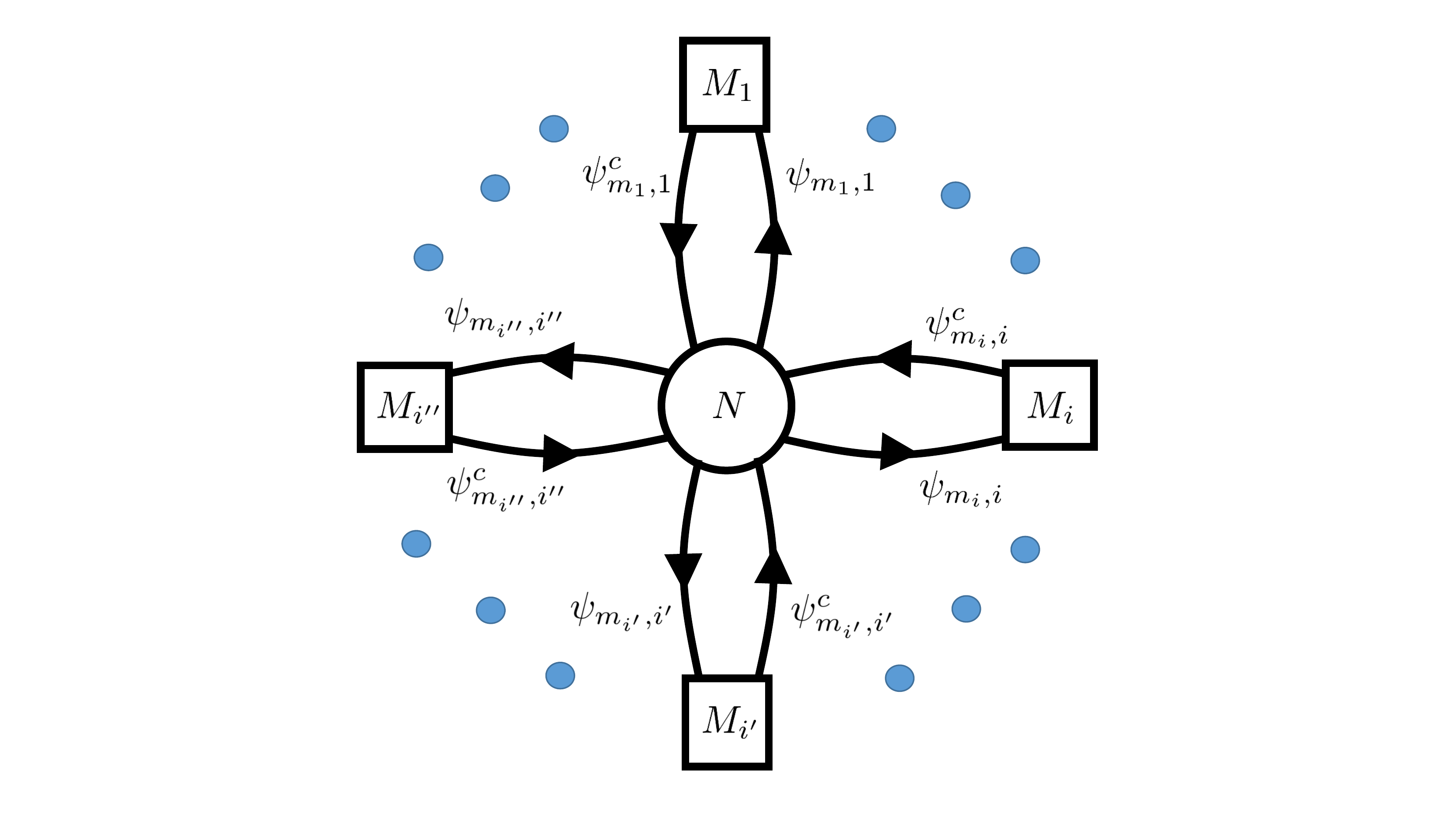}
\caption{Depiction of the star shaped quiver generated by an intersecting brane configuration in F-theory.
The central node corresponds to the contribution from the 7-brane wrapped over the gauge theory curve,
and the satellite nodes indicated as squares correspond to the flavor branes of the system. These intersect
the Hitchin system curve at points, and for $k$ such intersecting fields, there are regions in the
moduli space which are represented by higher poles in the Hitchin system Higgs field.
Each such satellite node corresponds to the location of a distinct marked point.}
\label{StarQuiver}%
\end{figure}

Another approach to the study of Hitchin systems with poles is the use of star
shaped quivers (see e.g. \cite{STAR,2016arXiv160403382H,2016arXiv160908226R} and associated work on hyperpolygons \cite{2011arXiv1101.3241G}). Mathematically, the connections between such quivers and wild Hitchin systems have been viewed as a novel correspondence between disparate geometric objects. Here, we would like to understand how this structure emerges naturally
from a physical point of view. Since we have an
$SU(N)$ gauge theory, we have a central quiver node with this gauge group. In
mathematical terms, we have a copy of the fundamental representation, namely the vector space
$\mathbb{C}^{N}$. Now, once we include the presence of
intersections with additional 7-branes, there are additional
bifundamental fields, the $\psi\oplus\psi^{c}$. For each stack of $M_{i}$
7-branes intersecting the Hitchin curve at a marked point $p_{i}$, we have
a $SU(M_{i})$ flavor symmetry, with defining representation a copy of
$\mathbb{C}^{M_{i}}$. As we have already remarked, the order of vanishing for
the parent Higgs field dictates the total number of such bifundamentals, so we
introduce an additional label as $\psi_{(1,i)}\oplus\psi_{(1,i)}^{c}%
,...,\psi_{(k_{i},i)}\oplus\psi_{(k_{i},i)}^{c}$. For each such pair, we get
maps:%
\begin{align}
\psi_{(s,i)}  & \in\text{Hom}(%
\mathbb{C}
^{N},\mathbb{C}^{M_{i}}),\\
\psi_{(s,i)}^{c}  & \in\text{Hom}(\mathbb{C}^{M_{i}},%
\mathbb{C}
^{N}).
\end{align}
See Figure \ref{StarQuiver} for a depiction of this quiver. We construct a higher order
pole for the Higgs field of the Hitchin system using suitable
bilinears in the $\psi$ and $\psi^{c}$'s. Referring to this pairing as before, namely
$\langle \langle \cdot , \cdot \rangle \rangle $, we have:
\begin{equation}
\Phi\sim du \underset{i}{\sum}\underset{l_{i}=1}{\overset{k_{i}}{\sum}}%
\frac{\underset{s+t=l_{i}}{\sum} \langle \langle \psi_{(s,i)}^{c} , \psi_{(t,i)} \rangle \rangle}{\left(  u-p_{i}\right)  ^{l_{i}}}+...
\end{equation}
where the ellipsis \textquotedblleft...\textquotedblright\ refers to regular terms. In this case, the notion of
Stokes chambers and Stokes data must be consistently combined across different patches.
We refer to Appendix \ref{app:wild} for some examples of this analysis.

\subsubsection{Coordinate Free Formulation}

In our physical derivation of wild Higgs fields, we made use of a particular
coordinate system with matter localized at the non-reduced scheme $u^k = 0$.
Indeed, as we have already remarked, one of the subtle features of wild Higgs fields
is the fact that the higher order poles are not detected by a residue formula.
In this subsection we develop a coordinate free formulation of the same data obtained above.

Consider the curve $C$ of genus $g$, with marked points $p_i$ with multiplicities
$n_i$. We view the $p_i$ as the locations of the poles for the Higgs field,
and the $n_i$ as the order of each pole. Given the effective divisor $D=\sum_i n_i p_i$, we consider the sheaf $\mathcal{O}_D$  of holomorphic functions on $D$ as a subscheme of
$C$, so that multiplicities are considered.  Since functions can be
multiplied, $\mathcal{O}_D$ is a sheaf of rings.
Then the ring
${\mathbb{C}[u]}/{\left(  \alpha
_{R}(u)\right)  }$ appearing in (\ref{localreps}) can be more intrinsically
written as $H^0(\mathcal{O}_D)$.  Another useful way to think of
$\mathcal{O}_D$ is as the quotient $\mathcal{O}_C$ by the ideal sheaf of functions
vanishing on $D$ (including multiplicities).  Since the ideal sheaf of $D$
is isomorphic to $\mathcal{O}_C(-D)$, we have a short exact sequence
\begin{equation}
  0\to \mathcal{O}_C(-D)\to \mathcal{O}_C \to \mathcal{O}_D\to 0.
\label{od}
\end{equation}

It turns out that there is a natural notion of differentials with poles on $D$
including multiplicity in terms of the notion of the dualizing sheaf
$\omega_D$ of $D$ \cite{HARTSHORNE}. This is a generalization to singular
schemes of the canonical bundle of a smooth variety.  Since $D$ is just
a collection of points, there is a simpler coordinate-free description
of $\omega_D$ in terms of 1-forms and poles (which we will also
derive below), but
we include the more general description of the dualizing sheaf
here in anticipation of applications
to 4D models in which we can have defects along singular curves.

The dualizing sheaf can be computed as in \cite{HARTSHORNE} by considering
\begin{equation}
  \label{duald}
  \omega_D=\underline{\mathrm{Ext}}^1_{\mathcal{O}_C}\left(\mathcal{O}_{D},\mathcal{O}(K_C)
\right),
\end{equation}
where $\omega_C=\mathcal{O}(K_C)$ is usual sheaf associated with the canonical
bundle of $C$ and $\underline{\mathrm{Ext}}^1_{\mathcal{O}_C}$
denotes the Ext sheaf, rather than the Ext group.  Applying the long exact
sequence of $\underline{\mathrm{Ext}}^*_{\mathcal{O}_C}(\cdot,\mathcal{O}(K_C))$ to (\ref{duald}) gives
\begin{equation}
  \label{extles}
  \underline{\mathrm{Ext}}^0_{\mathcal{O}_C}\left(\mathcal{O}_C,\mathcal{O}(K_C)\right) \to
  \underline{\mathrm{Ext}}^0_{\mathcal{O}_C}\left(\mathcal{O}_C(-D),\mathcal{O}(K_C)\right) \to
  \underline{\mathrm{Ext}}^1_{\mathcal{O}_C}\left(\mathcal{O}_D,\mathcal{O}(K_C)\right) \to
  \underline{\mathrm{Ext}}^1_{\mathcal{O}_C}\left(\mathcal{O}_C,\mathcal{O}(K_C)\right).
\end{equation}
The terms in (\ref{extles}) can all be identified since
$\underline{\mathrm{Ext}}_{\mathcal{O}_C}^0$ is just the $\mathrm{Hom}$ sheaf, while
\begin{equation}
\underline{\mathrm{Ext}}^1(\mathcal{O}_C,\mathcal{O}(K_C))=0,
\end{equation}
since $\mathcal{O}_C$ is locally free.  So (\ref{extles}) becomes
\begin{equation}
  \mathcal{O}(K_C)\to \mathcal{O}(K_C+D)\to\omega_D\to 0.
\label{extlesev}
\end{equation}
Moreover, the first map in (\ref{extlesev}) is just multiplication by the
local equations defining $D$, and thus we see that $\omega_D$ is just the restriction
of $\mathcal{O}(K_C+D)$ to $D$:
\begin{equation}
\omega_D=\mathcal{O}(K_C+D)|_D=:\mathcal{O}_D(K_C+D).
\label{omegad}
\end{equation}
At a point $p$ with local coordinate $u$ occurring with multiplicity $k$ in $D$,
then the sections of $\omega_D$ at $p$ are precisely the expressions
\begin{equation}
  \label{localomegad}
  du \left(\frac{T_k}{u^k}+\ldots+\frac{T_1}{u}\right)
\end{equation}
modulo regular terms, since we have modded out by the regular differentials
$\mathcal{O}(K_C)$ in (\ref{extlesev}).

Now $\omega_D$ is a sheaf of modules on $D$ rather than a sheaf of rings.
In down to earth terms, though we cannot multiply differentials, we can multiply
functions and differentials to get a differential.  Furthermore,
$\omega_D$ is a free sheaf and hence one has an isomorphism of sheaves of modules (not rings):
\begin{equation}
  \label{free}
\beta:\mathcal{O}_D\simeq\omega_D,
\end{equation}
 In this language the apparent coordinate
dependence in the wild system is the statement that $\beta$ is not intrinsic.
The ambiguity that we faced earlier in the choice of local
coordinates is replaced by an ambiguity in a choice of isomorphism
(\ref{free}), which occurs only at the points of $D$ rather than in a
neighborhood.  We can use coordinates to define an isomorphism by
\begin{equation}
  \label{isom}
T_k+\ldots T_1u^{k-1}\mapsto
  \left(\frac{T_k}{u^k}+\ldots+\frac{T_1}{u}\right)du,
\end{equation}
as an isomorphism which respects multiplication by sections of $\mathcal{O}_D$.
But any isomorphism of modules will do.

More abstractly, we can see (\ref{free}) without computation by noting
that $D$ is locally defined by one equation (e.g.\ by $\alpha$ in our
earlier notation) hence $D$ is Gorenstein, which implies that $\omega_D$
is locally free \cite{HARTSHORNE}.  Then since $D$ just consists of finitely
many points, the local isomorphism is in fact a global isomorphism.

From the above analysis, we can rewrite (\ref{localreps}) as\footnote{Here we drop the factors of
$K_{T^2}^{1/2}$ appearing in  (\ref{localreps}) since we are   dealing with the vacua of
a 6D theory.}
\begin{equation}
  \label{locrepint}
  \psi\in R\otimes H^0(\mathcal{O}_D),\qquad \psi^c\in R^c\otimes
H^0(\mathcal{O}_D).
\end{equation}
Then to get the F-term equation of motion, we combine multiplication in $\mathcal{O_D}$, the
represen-tation-theoretic pairing $\langle \langle\  \cdot , \cdot \rangle \rangle$, and the isomorphism
$\beta$ to define $\beta(\langle \langle \psi,\psi^c \rangle \rangle)$, an adjoint-valued section
of $\omega_D$, which is just the polar part of a 1 form at each point of $D$
as explained above.  So the equation of motion on the curve $C$ can be rewritten as:
\begin{equation}
  \label{eom}
  \bar\partial_A(\Phi)=0\ {\rm \ on\ }C-D,\ {\rm singular\ part\ of\ }
\Phi = \beta(\langle \langle \psi,\psi^c \rangle \rangle).
\end{equation}

We now check that the physics does not depend on the choice of $\beta$.  Given
a second isomorphism $\beta':\mathcal{O}_D\to\omega_D$, we exhibit an
explicit redefinition of the fields $\psi,\psi^c$ which takes the
equation of motion for $\beta$ to the equation of motion for $\beta'$.
Indeed, consider $\mathfrak{A}=(\beta')^{-1}\circ\beta$, an automorphism of the free module
$\mathcal{O}_D$.  Let $f=\mathfrak{A}(1)\in H^0(\mathcal{O}_D)$.  Then using the
module homomorphism property of $\mathfrak{A}$, we see that $\mathfrak{A}=m_f$,
the automorphism $m_f(g)=fg$ of $\mathcal{O}_D$ given as multiplication by
$f$.

Since $\mathfrak{A}$ is an automorphism, we have that $f(p_i)\ne0$ for each $p_i$.  So we can
find a well-defined square root of $f$ which we write as $\sqrt{f}\in
H^0(\mathcal{O}_D)$.  Then via the field redefinition for $\psi$  \begin{equation}
  \label{psiredef}
  R\otimes H^0({\mathcal{O}_D})\stackrel{1\otimes m_{\sqrt{f}}}\longrightarrow
R\otimes H^0({\mathcal{O}_D}),
\end{equation}
and a similar field redefinition for $\psi^c$, we verify that the
equation of motion for $\beta$ are transformed into  the equation of motion
for $\beta'$, since the combined effect of these transformations on
$\langle \langle \psi,\psi^c \rangle \rangle$ is just $1\otimes m_f$.
The treatment of bifundamental matter works in  the same fashion, in which case one
includes a sum over flavors in our definition of the pairing $\langle \langle \cdot , \cdot \rangle \rangle$.

\subsection{Moduli Spaces}

Now that we have presented a general method for constructing
Higgs fields with higher order poles, it is natural to ask
about the resulting moduli space of vacua.

To a certain extent, this depends on the physical context of the problem. If we treat the Hitchin system with wild ramification as the
full system, then we have a well-defined moduli space problem which has previously been studied in the math literature. We shall shortly review this parametrization of the moduli space, but first we would like to understand what sort of constraints embedding in F-theory or type IIA string theory imposes on these systems.

To illustrate some of these points, consider first the case of 6D F-theory vacua. For a 6D gauge theory with gauge group $G$,
there is typically a coupling to 6D tensor multiplets as well as 6D hypermultiplets. Anomaly cancellation for each such gauge
group factor imposes tight constraints on the total number of hypermultiplets which are present. For example, in the case of a local
base $\mathcal{O}(-2) \rightarrow \mathbb{P}^1$, consider an $SU(N)$ 7-brane wrapped on the $\mathbb{P}^1$. Cancellation of anomalies
then yields the condition that there are precisely $F = 2N$ matter fields in the fundamental representation of $SU(N)$. Similar considerations
hold for other curves and gauge group factors for 6D vacua.

Such constraints are seemingly less stringent in the related context of
4D $\mathcal{N} = 2$ vacua obtained, for example, by compactifying on a further $T^2$ factor. For example, as this is
not a chiral theory in four dimensions, the constraints from gauge anomaly cancellation are no longer present. There are
still constraints, however, because at least in a local model, we must require that we can write a complete metric
for the local Calabi-Yau in the neighborhood of the Hitchin system curve. Otherwise, we must include additional sectors and / or
reintroduce gravity into the model. If gravity is to remain
decoupled on a $-2$ curve engineering an $SU(N)$ gauge theory, we must
require that the number of 4D hypermultiplets in the fundamental representation
is $F \leq 2N$. So again, we see that physical constraints limit
the total number of matter fields.

But once we accept that the total number of matter fields is bounded above, we must also accept that the order of poles
in any wild Hitchin system will also be bounded above. Indeed, our whole method for generating higher order poles relies
on activating background values for these matter fields. To illustrate,
we see that for $k_i$ hypermultiplets transforming in the bifundamental
representation $(\mathbf{N} , \overline{\mathbf{M}_i})$, the total number of flavors $F_{\mathbf{N}}$ in the fundamental representation is:
\begin{equation}
F_{\mathbf{N}}=\underset{i}{\sum}k_{i}M_{i}.
\end{equation}

Neglecting the singlet moduli (which we associate with a decoupled gravitational sector),
the moduli space swept out by these charged fields are associated
with the bulk Higgs field, the gauge connection, modulo complexified gauge
transformations, for a total of $(2g-2)\dim G$ complex degrees of freedom, as well as
$F_R$ localized matter fields in some representation $R$ of the gauge group.
The dimension of the moduli space from these 7-branes is then:
\begin{equation}
\dim\mathcal{M}_{\text{7-branes}}=(2g-2)\dim G + \underset{R}{\sum} 2 F_R \times\dim R,
\end{equation}
in the obvious notation.

Let us now match this structure to the moduli space of the wild Hitchin system.
As we have already remarked, some care must be exercised in even defining this moduli space, since we are dealing
with a singular field configuration. For simplicity, consider the moduli space on a genus $g$ curve with a single marked
point at $u = 0$ in which the Higgs field has singular part:
\begin{equation}
\Phi = du \left(\frac{T_k}{u^k} + ... + \frac{T_1}{u} \right).
\end{equation}

Given some fixed choice of the $T_{i}$'s one can consider solutions to the Hitchin
system. This question is studied in \cite{BoalchHabil, Witten:2007td, BoalchIso, BacardiCola}
for the case where all $T_{j}$ are regular and semi-simple (that is, diagonalizable
with all eigenvalues non-zero). Call this moduli space $\mathcal{M}_{H}%
(T_{1},...,T_{k})$. One can also consider the moduli space associated where
$T_{1}$ is not held fixed, which we denote by $\mathcal{M}_{H}(T_{2}%
,...,T_{k})$. In reference \cite{Witten:2007td}, the former is denoted by $\mathcal{M}%
_{H}(T_{1})$ while the latter is denoted by $\mathcal{M}_{H}$. This notation
emphasizes the point that the complex structure and symplectic structure of
the moduli space does not depend on the higher order polar terms. To avoid
confusion, we shall keep manifest all $k$ terms in what follows.

The dimension of the moduli space is calculated in \cite{Witten:2007td, BacardiCola},
with the end result:%
\begin{equation}
\dim\mathcal{M}_{H}(T_{1},...,T_{k})  = (2g-2)\dim G+k(\dim G-\text{rk}G).
\end{equation}
The contribution $(2g-2)\dim G$ is the dimension of the Hitchin
moduli space on a genus $g$ curve, and the additional
terms are associated with a single order $k$ pole.
From the parameters $T_{1},...,T_{k}$, we have $k\times\dim G$ complex
parameters. Of these, a topological interpretation either from a holonomy or
topological Stokes data can be given to $k\times\dim G-(k-1)\times$ rk$G$ of
these parameters. The remaining $(k-1)\times$ rk$G$ parameters are then
associated with isomonodromic deformations \cite{JIMBO}. See Appendix \ref{app:wild} for further
discussion as well as some explicit examples.

Let us now determine the dimension of the moduli space in accord
with an algebro-geometric construction, geared towards an eventual F-theory
construction. We calculate the contribution to the various Casimir invariants
which are independent of regular terms. Focusing on the special case $G=SU(N)$, we have:%
\begin{equation}
\text{Tr}(\Phi^{j})=du^{\otimes j} \left(\frac{\text{Tr}(T_{k}^{j})}{u^{jk}}+...+\frac
{j\text{Tr}(T_{k}^{j-1}T_{1}+...)}{u^{k(j-1)+1}}+\text{(contributions from
regular terms)}\right).%
\label{irregterms}
\end{equation}
Following \cite{Witten:2007td}, expressions of the form (\ref{irregterms}) form an
affine space isomorphic to
\begin{equation}
H^{0}(C,K_{C}^{j}\otimes \mathcal{O}(p)^{k(j-1)}),
\end{equation}
since any two expressions of the form (\ref{irregterms}) differ by a
$j$-differential with a pole at $p$ of order at most $k(j-1)$.

Let us count the dimension of this moduli space. From Riemann-Roch, we have:%
\begin{equation}
\dim H^{0}(C,K_{C}^{j}\otimes \mathcal{O}(p)^{k(j-1)})=\left(
j-\frac{1}{2}\right)  (2g-2)+k(j-1).
\end{equation}
Summing over the independent Casimirs, i.e. from $j=2$ to $j=N$, we have the
total number of such deformation moduli is:%
\begin{equation}
\dim\mathcal{M}_{\text{cplx}}=\underset{j=2}{\overset{N}{\sum}}\dim
H^{0}(C,K_{C}^{j}\otimes \mathcal{O}(p)^{k(j-1)})=\frac{1}{2}\left[
(2g-2)\dim G+k(\dim G-\text{rk}G)\right]  ,
\end{equation}
where in the above we used the fact that $\dim G=N^{2}-1$ and rk$G=N-1$. Now,
in addition to the complex structure moduli captured by the base of the Hitchin system,
we also have (in the IIA\ formulation) the intermediate
Jacobian of the Calabi-Yau threefold defined by the local model.
In the Hitchin system formulation, these originate from the Prym variety
of the spectral curve \cite{HitchinSelf, Beauville}.
The total dimension of the moduli space is therefore:%
\begin{equation}
\dim\mathcal{M}_{H}(T_{1},...,T_{k})=(2g-2)\dim G+k(\dim G-\text{rk}G).
\label{Mtot}%
\end{equation}

In these calculations, we have  used the fact that the relevant moduli can be counted
solely from the Casimirs of the Higgs field without regard to the Higgs field itself.
This follows from the fibration structure of the integrable system reviewed in
Appendix \ref{app:iscft}.

\section{Geometric Unification \label{sec:GLOBAL}}

Our discussion in the previous sections focused on the physical origin of wild Hitchin systems
in local F-theory models. To complete the circle of ideas we now present some illustrative examples
generated by successive tuning in the limiting behavior of complex structure moduli.
We demonstrate that this global perspective also unifies different wild Hitchin
systems in one geometric framework.

In a global F-theory model the complex structure moduli can either be localized on a component of
the discriminant locus -- as in the case of charged matter -- or can be moduli which transform as singlets under
all gauge group factors. In the latter case, such moduli can often be thought of as remnants from Higgsing a
higher rank gauge group. This often occurs in the unfolding of colliding singularities. For example, if we have matter fields in a
representation $R_{\mathrm{parent}}$ of a parent gauge group $G_{\mathrm{parent}}$, decomposition into irreducible representations of a
possibly semi-simple descendant gauge group $G$ can include singlets of any or all of the simple gauge group factors of $G$.

There are also ``purely gravitational'' gauge singlet hypermultiplets
which should best be viewed as moving the locations of various marked points.
Geometrically, these define torsion deformations of the local model.
In Appendix \ref{app:cplx} we extend the analysis presented in \cite{Anderson:2013rka},
showing that all such deformations are physical moduli in an F-theory model.

Including such closed string sectors provides a unified perspective for wild systems. For example, since we can move the locations
of various poles, we can address what happens when different marked points with possibly different matter content
and pole orders are brought together at a single location in the geometry.
This question is difficult to answer in the wild Higgs bundle literature, because even having wild ramification
at different points requires introducing additional
gluing conditions known as tentacles which match the Stokes chambers present near distinct marked
points. Additionally, there is always the possibility that additional moduli must be incorporated into the structure
of the wild system to properly account for such gluing operations. From the geometric perspective,
however, all such moduli spaces are on an equal footing and simply correspond to different parameterizations
specified by the singlet moduli of the global F-theory model.

To illustrate, consider an $SU(2)$ 7-brane gauge theory localized
on a $\mathbb{P}^1$ with self-intersection $-2$ in an F-theory base. As we have already remarked,
6D anomaly cancellation requires precisely four matter fields in the fundamental representation of $SU(2)$. These matter fields
may be localized at distinct points of the curve, or could be collected together at the same point. As we do this,
the pole structure, as well as flavor 7-branes of the system will also change. The moduli controlling these
deformations are those responsible for moving the locations of marked points. As we establish in Appendix \ref{app:cplx}, such moduli are
part of the physical moduli of an F-theory model.

To provide some further examples, in this section we will focus
on F-theory compactified on an elliptically fibered Calabi-Yau threefold
$X \rightarrow \mathbb{F}_n$ with base a Hirzebruch surface.
Consistency of the model requires $-12 \leq n \leq 12$.
The minimal Weierstrass model for such geometries is:
\begin{equation}
y^2 = x^3 + fx + g
\end{equation}
with:
\begin{align}
f(w,u)  &  \sim \underset{i}{\sum}w^{i}f_{8+n(4-i)}(u)\\
g(w,u)  &  \sim \underset{j}{\sum}w^{j}g_{12+n(6-j)}(u)
\end{align}
where $i$ is bounded by the largest value less than or equal to $8$ such that $8+n(4-I) \geq 0$ and $j$ is bounded by the largest number
less than or equal to $12$ with $12 +n(6-J) \geq 0$. Here $w=0$ determines
the zero section of the Hirzebruch surface (i.e. the base $\mathbb{P}^1$) and $u$
is a local coordinate on the base $\mathbb{P}^1$. By tuning the Weierstrass coefficients, namely by adjusting the
values of charged and gauge singlet hypermultiplets, we will show how to interpolate between various types of wild Hitchin systems.

Our plan in the rest of this section will be to present
some illustrative examples, focusing on models
where we preserve an $SU(2)$ gauge group factor. After this, we
revisit the analysis presented in \cite{Anderson:2013rka} for the F-theory dual to
the Heterotic tangent bundle on a $K3$ surface.

\subsection{A Wild Compact $SU(2)$ Model}

To illustrate the above points, let us consider an explicit example.
We consider an F-theory model with base
$\mathbb{F}_{n}$ for $n$ a non-negative integer.
We will be interested in the case where the $w=0$ locus supports an
$\mathfrak{su}(4)$ gauge symmetry, which is associated with an $I_{4}$ fiber.
Recall that this means $f$ and $g$ must not vanish along $w=0$ but the
discriminant must vanish to order $w^{4}$.  Furthermore, we will require
that this locus intersects a curve supporting an $\mathfrak{su}(2)$ symmetry (i.e. $I_2$ fiber).
Producing a product group requires tuning singularities on two loci in
the base, $w=0$ and $w=\epsilon$ simultaneously, which is difficult to
identify when $f$ and $g$ are expanded in $w$ alone. This tuning
process was described in \cite{Anderson:2015cqy}, where $f$ and $g$
can be systematically expanded in $w(w-\epsilon)$. Letting $\sigma= w - \epsilon$ for brevity:
\begin{align}\label{double_expand}
f & = F_0 + F_1 w\sigma +  F_2 w^2\sigma^2 + \ldots \\
g &= G_0 + G_1 w\sigma +  G_1 w^2\sigma^2 + \ldots ,
\end{align}
where $F_i = f_{2i} + f_{2i+1}u$ and $G_i = g_{2i} + g_{2i+1}w$.
For this example, we will set $\epsilon=\epsilon_1 \beta$
(note that in the limit $\epsilon_1 \to 0$ this leads to an $SU(6)$ theory).

It is convenient to define the $\mathfrak{su}(2)$ locus as
$\sigma=w-\beta \epsilon_1$ where $\beta$ is a polynomial of degree
$r$ and $\epsilon_1$ has degree $n-r$ over the $\mathbb{P}^1$ base.
To leading order the Weierstrass coefficients take the form
\begin{align}
f &=  -\frac{\alpha ^4 \beta ^4}{48}+\frac{1}{18}  \left(2 \alpha ^2 \beta ^2 \text{$\epsilon_1$} \text{$\phi$}-3 \alpha ^2 \beta ^3 \nu \right)w + \ldots \\
g &=  \frac{\alpha ^6 \beta ^6}{864}+\frac{1}{216}  \left(3 \alpha ^4 \beta ^5 \nu -2 \alpha ^4 \beta ^4 \text{$\epsilon_1$} \text{$\phi$}\right)w + \ldots
\end{align}
with corresponding discriminant locus
\begin{align}\label{delta}
\Delta=& (\sigma)^2\left(w^4\right)\left(\frac{1}{5184}(\alpha ^4 \beta ^2 )[12 \beta  \text{$\phi$}^3 (\alpha ^2+2 \nu  \text{$\epsilon_1$})+ \ldots] +{\cal O}(w)+\ldots \right)
\end{align}
where in addition to the functions $\beta$ and $\epsilon_1$, the solution is parameterized by functions $\alpha$ (of degree $2+n-r$), $\phi$ (of degree $4+r$), and $\nu$ (of degree $4+n-r$).

The total matter content\footnote{The singlet moduli counted above are only those contributing on the patch containing $w=0$. The degrees of freedom in $f_i$ with $i\geq 4$ and $g_j$ with $j \geq 6$ are unconstrained and omitted from the count above. In the heterotic dual theory these degrees of freedom are merely the 20 moduli of the heterotic $K3$ and the $-30n+112$ moduli associated to an $E_8$ bundle with $c_2=12-n$.} of the theory is determined by the two integers ($n$ and $r$) and is given in Table \ref{table1}. It is clear that $\epsilon_1=0$ counts the bifundamental matter, while $\alpha=0$ corresponds to the ${\bf (6,1)}$. Intersections with the $I_1$ component of the discriminant gives $2n+16+2r$ ${\bf (4,1)}$ multiplets and $2n+16+r$ ${\bf (2,1)}$ multiplets. The locus with $\beta=0$ corresponds to a $D_5$ enhancement which gives $\frac{r}{2}$ ${\bf (6,2)}$ multiplets. Observe that especially for the $SU(4)$ gauge theory, we have matter fields in different representations. This can also be covered in a local model, and the Hitchin system for such a case has recently been studied for example in reference \cite{BradlowSchaposnik}.
\begin{table}[t!]
\begin{center}
{\begin{tabular}{@{}cccc@{}}
\hline
Representation & Multiplicity&Representation & Multiplicity \\ \hline
$({\bf 1},{\bf 1})$ &  $4n-2r+22$&$({\bf 4},{\bf 1})$ & $2n+2r+16$\\
$({\bf 6},{\bf 2})$ & $\frac{r}{2}$&$({\bf 4},{\bf 2})$& $n-r$\\
$({\bf 1},{\bf 2})$ & $2n+r+16$&$({\bf 6},{\bf 1})$& $n-r+2$\\
\hline
\end{tabular}
\caption{{\it The multiplicity of matter fields (in full hypermultiplets) in the SU(4) $\times$ SU(2) theory on base $\mathbb{F}_n$.}}\label{table1}}
\end{center}
\end{table}
It is illuminating to consider the degrees of freedom visible \emph{only} on the $SU(2)$ component of the discriminant locus in the case that there are no ${\bf (6,2)}$ anti-symmetric fields. In this case $r=0$ and we have $n$ bifundamentals and $2n+16$ fundamentals of $SU(2)$. There are $4n+22$ total moduli of the system, plus $3n+18$ additional $SU(2)$ singlets not visible from the $SU(2)$ component of the discriminant (i.e. matter in the ${\bf (4,1)}$ and ${\bf (6,1)}$ representations). The previously mentioned example of a local $-2$ curve with $SU(2)$ gauge theory corresponds to the special case where we take $n = -2$ and $r = 0$. Note that in this case, we cannot retain the $SU(4)$ enhancement locus. Indeed, otherwise some of the entries in Table \ref{table1} would have negative multiplicity.

Let us now turn to the Hitchin system interpretation of this model. Assuming we have tuned
the moduli of the Weierstrass model to have an $SU(4) \times SU(2)$ gauge theory, we see that we actually have two Hitchin systems which are coupled via the source terms provided by the localized matter. Additionally, we see that generically, different sorts of bifundamental representations will be present. Now, from the perspective of the parent gauge theory described in section \ref{sec:WILD}, we also observe that poles in the Higgs field require background values for a single hypermultiplet.
In other words, even if we try to tune the moduli so that different matter fields localize at the same point of a curve, there is no pairing available between different matter field representations. So in this sense, these tunings of different representations cannot change the pole order, but only the location of non-zero entries for each generalized residue.

Assuming that all such localized matter are kept at distinct points, we see that from the perspective of the $SU(2)$ factor, we can engineer only simple poles. Even so, it should still be noted that in this case it is nonetheless possible to achieve a generic T-brane configuration for the $SU(2)$ factor. Conversely, we also see that for the $SU(4)$ factor, the presence of different representations, such as the $\mathbf{6}$ and $\mathbf{4}$, and the respective multiplicities allows us to fill out T-branes with simple poles.

Suppose next that we start to bring the various matter fields to the same location in the geometry. In this context it is possible to begin to engineer general wild T-brane solutions, embedded in $SU(2) \times SU(4)$. Using the techniques of section \ref{Sec:nucleation} to embed $SU(2)$ Hitchin systems into large gauge groups, we see that depending on the choice of structure group of the Higgs bundle, a variety of breaking patterns are possible, including unbroken 6D symmetry $SU(2) \times SU(2)$, $SU(4)$, and $SU(2)$ (as well as the maximally broken/unbroken cases).

We will consider several choices of Hitchin system which break some of the gauge symmetry of the $SU(4) \times SU(2)$ system. For ease of exposition, we shall first consider embedding a Higgs bundle into the $SU(2)$ factor. Given a choice of $n$ and $r$, we would like to know, for example, the maximal pole order available in a wild $SU(2)$ Hitchin system engineered in such a global model. If we choose to also break the $SU(4)$ gauge symmetry (i.e. break the full gauge group), we can now expect that doublets of the $SU(2)$ Higgs bundle which descend from different sources can all combine
to produce a higher order pole. The total number of such doublets is:
\begin{equation}\label{firstdoubcount}
F_{\mathrm{doub}} = 3r + (2n + r + 16) + 4n - 4r = 6n + 16.
\end{equation}
(note that as expected by anomaly cancellation, this is the same number of doublets arising in the 6D theory from a single, generic $SU(2)$ symmetry on $\mathbb{F}_n$ \cite{BershadskyPLUS}). From this, we conclude that the highest order pole which can be achieved in such models is obtained by a maximal breaking pattern
of the $SU(4)$ factor, and with $SU(2)$ Higgs bundle pole order:
\begin{equation}
k_{\mathrm{max}} = 6n + 16.
\label{eq:poleorder}
\end{equation}
As a check, we will arrive at (\ref{eq:poleorder}) in Appendix~\ref{app:cplx} by geometric methods. For example, when $n =-2$, we can achieve a single pole of order $4$. We study the moduli space of wild Hitchin systems for this case in much
greater detail in Appendix \ref{app:wild}. At the absolute extreme where $n = 12$,
we can achieve a pole of order $88$.

Now, in contrast to the case above, let us consider embedding a Higgs bundle with structure group $SU(2) \subset SU(4)$ only. This choice leaves unbroken $SU(2) \times SU(2)$ in the 6D theory and gives rise to an anomaly consistent spectrum of $n$ ${\bf(2,2)}$ bifundamentals and $4n+16$ ${\bf [(1,2)+(2,1)]}$ representations. Here then the relevant counting of doublets ``visible" to the $SU(2)$ Higgs bundle is obtained by the branching rules of $SU(4) \to SU(2)$ and the counting the original number of ${\bf 4}$ and ${\bf 6}$ multiplets in the CY geometry, leading to
\beq
F'_{\mathrm{doub}} = r +(n-r+2) + (2n + 2r + 16) + 2(n-r) = 5n + 18.
\eeq
In this case the maximal pole order is clearly different than the case above. For example, if only the location of the ${\bf(6,2)}$ and ${\bf(4,2)}$ multiplets are forced to overlap, then the maximal pole order in the $SU(2) \subset SU(4)$ Higgs bundle is  determined by $2n-r= 2k_{\mathrm{max}}$. In the case that $r=0$ we have the much more restrictive bound of $k_{\mathrm{max}}=n$. We provide these examples to illustrate that the for the \emph{same CY geometry} there can be multiple choices of T-brane solutions, with different wild pole structure, leading to entirely different 6D effective theories.

In Appendix~\ref{app:cplx}, we will
study bifundamentals associated with $SU(M)$-$SU(N)$ collisions from a
geometric perspective.  While we do not explicitly study collisions with
additional matter as in the situation of Table~\ref{table1}, more general
situations can be understood in principle by combining the analyses of
bifundamentals and fundamentals considered separately in Appendix~\ref{app:cplx}.

\subsection{Tangent Bundle to $K3$ Revisited \label{ssec:TAN}}

Most of the examples encountered so far in this paper have used an explicit weakly coupled hypermultiplet.
Recent work on 6D superconformal field theories has also shown that the correspondence between matter and complex structure
deformations also extends to the case of conformal matter as in references
\cite{DelZotto:2014hpa, Heckman:2014qba, Heckman:2016ssk, Mekareeya:2016yal}. Here we study this phenomenon
in one particularly tractable case: small instantons of heterotic theory which have been dissolved back into finite
size instantons. We focus on the case of deformations to an unbroken $E_7$ gauge group, namely, the breaking
pattern involves an $SU(2)$ gauge theory. Our plan in this subsection will be to study this case in greater detail,
following the analysis presented in reference \cite{Anderson:2013rka} (see also \cite{BershadskyPLUS, MorrisonVafaII}).

In the spectral cover construction, or equivalently in the F-theory dual realization,
we have a Hitchin system on a $\mathbb{P}^{1}$ with $12+n$ marked points, the
number of instantons in the heterotic dual description. In the theory with a smoothing
deformation, we have an $\mathfrak{e}_{7}$ singularity which degenerates to an
$\mathfrak{e}_{8}$ singularity at marked points. The minimal Weierstrass model
is:%
\begin{equation}
y^{2}=x^{3}+\varepsilon f_{8+n}(u)w^{3}x+g_{12+n}(u)w^{5},
\end{equation}
that is, we have $8+n$ half hypermultiplets in the $\mathbf{56}$ of $E_{7}$.
The zeroes of $g_{12+n}(u)$ indicates the locations of the small instantons.
Dissolving these small instantons into flux amounts to activating a source for
the Higgs field, triggering a breaking pattern down to $E_{7}$. The parameter
$\varepsilon$ indicates that we can also take a limit where we proceed back to
a singular T-brane configuration.

We review and extend the description from \cite{Anderson:2013rka}
of this breaking pattern in terms of Hitchin systems.
Let $D\subset\mathbb{P}^{1}$ be the divisor
consisting of the $12+n$ zeros of $g_{12+n}$.
There is essentially no change in our conclusions if some of
the zeros of $g_{12+n}$ have multiplicity greater than one.

The breaking pattern from $E_8$ to $E_7$ is governed by an $SU(2)$ Hitchin system.
So, we consider meromorphic $SU(2)$ Higgs bundles
\begin{equation}
  \Phi:E\to E\otimes K_{\mathbb{P}^1}(D)
\end{equation}
with simple poles on $D$.  Here, if $p$ occurs in $D$ with multiplicity $m$,
by a ``simple pole on $D$'' we mean that the poles of $\Phi$ at $p$ can have
order $m$.  Our conclusion is that $\mathrm{Tr}(\Phi^2)$ also has simple
poles on $D$, just as in the situation where the zeros of $g_{12+n}$ are
isolated.

Next, consider the parameter space
\begin{equation}
  B=H^0(\mathbb{P}^1,K^{2}_{\mathbb{P}^1}(D))
\end{equation}
Note that $\dim B=n+9$ when $m=n+12$. For each $b\in B$,
we construct a spectral cover $C_b$ of $\mathbb{P}^1$ inside
the total space of the line bundle $K_{\mathbb{P}^1}(D)$.
Letting
$y$ be a point of this total space, then the spectral cover $\pi:C_b
\to \mathbb{P}^1$ is given
by the equation
\begin{equation}
  y^2=g_{n+12}b.
\label{eq:spectralcover}
\end{equation}
Note that $g_{n+12}b\in H^0(\mathbb{P}^1,K_{\mathbb{P}^1}^2(2D))$, so
(\ref{eq:spectralcover}) makes sense.

If we have a line bundle $L$ on $C_b$ (or torsion-free sheaf more
generally), we recover a meromorphic Higgs bundle by putting
$E=\pi_*L$.  The embedding of $C_b$ in the total space of
$K_{\mathbb{P}^1}(D)$ gives rise to
$\Phi:E\to E\otimes K_{\mathbb{P}^1}(D)$ in the same manner as
the case of a Higgs field without singularities.

Now $C_b$ is singular over the points $p\in \mathbb{P}^1$ at which
$g_{n+12}$ vanishes to higher order.
The important point for us is that these singularities do not affect the
moduli count.  This is because the tangent space to the moduli space
$H^1(C_b,\mathcal{O}_{C_b}^*)$ of line bundles on $C_b$ is
$H^1(C_b,\mathcal{O}_{C_b})$.  Since $C_b$ is connected, the dimension of
this vector space is just the arithmetic genus of $C_b$, which is a deformation
invariant so can be computed by Riemann-Hurwitz by assuming that the zeros
of $g_{n+12}$ are distinct.  The result is that the genus, hence the dimension,
is $n+9$.

There are $n+12$ additional pairs of moduli.  The moduli of $B$ can be enlarged
by letting the points of $D$ vary, i.e.\ the base space can be enlarged to
a fibration over
$Sym^{n+12}\mathbb{P}^1$ with typical fiber $B$.  The other half of the pair
of $n+12$ moduli appear in what are identified with RR moduli in the IIA
description, arising from the limiting mixed Hodge structure.

\section{Conclusions \label{sec:CONC}}

A central pillar of F-theory is the close
correspondence between the geometry of elliptically fibered Calabi-Yau
manifolds, and the resulting moduli space of vacua for the low energy
effective field theory. Vacua with T-branes generalize this correspondence
since they involve non-abelian intersections of 7-branes which are not
visible in complex geometry. In this paper we have
shown that the open string patch of this moduli space is governed by a Hitchin system coupled
to defects. Non-zero background values for these defects
provide a systematic way to build up T-branes localized at points as well as non-reduced schemes.
This gives a systematic method for engineering wild Higgs bundles in F-theory.
Calabi-Yau geometry also unifies seemingly different wild Hitchin systems:
changes in the order of poles or in the nature of residues are captured by corresponding variations in the
gauge singlet complex structure moduli of an F-theory compactification. Constraints from 6D effective field theory translate
to geometric conditions on the order of poles, and the rank of generalized residues which can actually be
realized by an F-theory model. In the remainder of this section
we discuss some avenues for future investigation.

An implicit feature of the results in this paper is the way in which limiting mixed Hodge structure is completed
in singular limits (as in \eqref{emergent_hitchin}) by an emergent Hitchin system \cite{Anderson:2013rka}. It would be instructive to develop
this explicit correspondence in more detail, both as a general ``proof of principle'' as well as a means to
further unify a priori distinct wild Hitchin systems.

Along these lines, wild Hitchin systems in F-theory must incorporate a sufficient number of
charged matter fields to realize certain types of higher order poles.
A natural question is whether there is a universal maximum upper bound possible in global models.
Using the results of this paper, we see that this question translates into determining a sharp upper bound on the
total amount of matter which can be charged under any particular gauge group in an F-theory model.

Another feature of wild Higgs bundles is the necessity of introducing Stokes chambers in order to properly account
for the generalized monodromy experienced by holomorphic sections of such singular bundles. Returning to the geometry
of an F-theory compactification, we expect that upon further compactification on a $T^2$ that such holomorphic sections will
correspond to matter fields localized at a point of the $T^2$ factor and spread over the Hitchin system curve in question. Since
such matter fields also correspond to localized deformations of the complex structure moduli of the Calabi-Yau, this also
suggests that the notion of Stokes chambers and generalized monodromy matrices should also lift to the complex structure moduli and intermediate Jacobian of the Calabi-Yau threefold. Developing the resulting effect on the periods of holomorphic three-forms with higher order singularities
-- namely not just simple poles along a divisor -- would be most instructive.

Now that we have a clear interpretation of most Hitchin system phenomena in terms of their corresponding F-theory
avatars, it is natural to consider next the resulting 4D vacua obtained from compactification on a Calabi-Yau fourfold.
The reduced supersymmetry means that we should not expect the moduli space of vacua to define a hyperkahler manifold.
Nevertheless, we expect the methods developed here and in \cite{Anderson:2013rka} to persist
in this more general setting.

Finally, one of the important phenomenological applications of T-branes is in the
construction of realistic Yukawa couplings. To extract these couplings,
it is necessary to know the localized profile of matter field wave functions. We have also seen that
the presence of higher order singularity types has an impact on the local presentation of matter fields,
since the actual normalizable wave functions depend on Stokes data. It would be interesting
to study the possible impact on wave function profiles, as well as the overlap of multiple wave functions.

\section*{Acknowledgements}

We thank P. Boalch, S. Gukov, R. Donagi, J. Hurtubise, D.R. Morrison,
S. Rayan, and W. Taylor for helpful discussions.
JJH thanks the 2015 Simons Center Summer Workshop on Geometry and Physics for kind
hospitality during part of this work. SK thanks UNC Chapel Hill
and the organizers of the Spring 2015 southeast conference
on string theory for kind hospitality during the completion of part of this work.
LBA, JJH, SK and LPS thank the organizers of the 2015 FRG\ workshop at Harvard for providing a stimulating
atmosphere for collaboration. The work of LBA is supported
by NSF grant PHY-1417337 and is a part of the working group activities of the 4-VA initiative ``A Synthesis of Two Approaches to String Phenomenology." The work of JJH is supported by NSF CAREER grant PHY-1452037.
JJH also acknowledges support from the Bahnson Fund at UNC Chapel Hill.
The work of SK was supported by NSF grants DMS-1201089 and
DMS-1502170. The work of LPS is supported by NSF grant DMS 1509693, and she would
like to thank the hospitality of IMPA, Brazil, and the support of the
GEAR Network through NSF grants DMS 1107452, 1107263, 1107367.

\newpage

\appendix

\section{Introduction to Wild Hitchin Systems}\label{app:wild}

In this Appendix we present some of the salient features of wild Higgs bundles.
Our aim here will be to give a self-contained introduction to those aspects relevant
for our discussion of wild T-branes in F-theory. We shall be interested in the Hitchin system on a genus $g$ curve $C$, for some gauge group $G$ (see \cite{2013arXiv1301.1981S,hitching2,Hitchin:1987mz} and references therein for relevant background material). In the physics literature, it is common to reference a gauge group $G$, an adjoint valued $(1,0)$ form $\Phi$, and a gauge connection $A$. It is convenient to instead work in terms of the pair $(E,\Phi)$, where $E$ is a stable holomorphic vector bundle with structure group $G$.

In order to describe the basic geometric features in the study of wild Higgs bundles, it is enough to  limit ourselves to Higgs bundles $(E,\Phi)$ over the projective line $\PP$. This is an effective divisor $D$, a vector bundle $\CE$ of rank $n$;
and an  $K_{\PP}(D)$ valued endomorphism  $\Theta:\CE\rightarrow \CE \otimes K_{\PP}(D)$
called the Higgs field. The divisor $D$ controls where $\Phi$ is allowed to have poles,
so it is sometimes referred to as the {\it polar divisor}.

It is useful to suppose that there is a compatible parabolic structure on $p_j$, in the sense of Mehta and Seshadri. This means, a set of {\it parabolic weights} which are  real numbers
\begin{eqnarray}
0\leq \alpha_0^j\leq\ldots \leq \alpha_{r-1}^{j}< 1,
\end{eqnarray}
and a finite decreasing filtration
$\{0\}=(\CE_{p_j})_1 \subseteq (\CE_{p_j})_{\alpha_{r-1}^j}\subseteq\ldots
\subseteq(\CE_{p_j})_0=\CE_{p_j}
$ of the fibre of $\CE$ at $p_j$ preserved by the residue ${\rm res}(\Phi,p_j)$, where $r_j$ is the smallest index  such that $\alpha^j_{r_j}>0$  (when such an index exists; $r_j=r$ otherwise). In what follows, we shall consider particular examples to illustrate the theory that has been developed in the literature to study Higgs bundles with poles.

We adopt the notation of reference \cite{BoalchAnnal} where a wild Higgs bundle of type $(m, r_1 , r_2 , \ldots)$ is a Higgs bundle with $m$ poles of orders $k_i:=r_i+1$. Then, as explained in \cite[Remark 9.12]{BoalchAnnal} for rank two Higgs bundles, one obtains moduli spaces of complex dimension two when
 $$(m, r_1 , r_2 , \ldots)=(4, 0, 0, 0, 0), (3, 1, 0, 0), (2, 1, 1), (2,
2, 0), (1, 3),$$
 denoting four poles of order 1;   two of order 2 and two poles of order 1; or two poles of order 2; or two poles of of order 3 and a simple pole; or finally one pole of order 4.

 \subsubsection*{A comment on notation} In the paper, and in particular in section  \ref{sec:WILD}, when considering Higgs bundles with a single pole of order $k$ at zero, it is convenient to adopt similar  notation to that used in \cite[Eq. (1.1)]{Witten:2007td} where the corresponding complexified  connection is written as
\begin{eqnarray}
\mathcal{A}=  du \left(\frac{T_k}{u^k}+\ldots+\frac{T_2}{u^2}+\frac{T_1}{u}\right),\label{L_wit1}
\end{eqnarray}
where  $T_1, T_2, \ldots , T_k$ are elements of the Lie algebra of group. On the other hand, in this appendix we would like to review some known results for low rank Higgs bundles (see e.g. \cite{BoalchIso}). To this end, we now write a complexified connection as:
\begin{eqnarray}
\mathcal{A}= d Q +T_1 \frac{du}{u},\label{par1}
\end{eqnarray}
for $Q$ a square matrix of meromorphic connections. From the above papers, one has the following definition.
\begin{definition}\label{mono11}
The {\bf formal monodromy} of $\mathcal{A}$ is
$M_0:=e^{2\pi i T_1}.$
Moreover, we say that $T_1$ is the {\bf exponent} of the formal monodromy.
\end{definition}
One should keep in mind that varying $Q$ from the perspective of reference \cite{BoalchIso},
will be considered equivalent to varying $(T_k,T_{k-1}, \ldots, T_2)$ from the perspective \cite{Witten:2007td}.

In what follows, we shall give a detailed description of data associated to Wild Higgs bundles of type $(1,1)$, and build up to two of the cases where the moduli space has dimension 2: the ones of $(1,3)$ and $(2,1,1)$ following closely \cite{PBsandwich} and references therein.

When calculating moduli dimensions, as done as in \cite[Remark 9.12]{BoalchAnnal} and references therein,  one needs to consider isomonodromic deformation \cite{JIMBO}, given by meromorphic differential
equations by which one can vary the parameters contained in $(T_k,T_{k-1}, \ldots, T_2)$  without
changing the generalized monodromy $\widehat{M}$ obtained as the product of the formal monodromy and the Stokes matrices (e.g. see \cite[Eq. (2.29)]{Witten:2007td}).

 \subsection*{Rank 2 Higgs bundles with 1 pole of order 2}\label{one2}
 We will follow \cite{BoalchIso} to study  Higgs bundles $(E,\Phi)$ on $\PP$ of rank $r=2$ and with only one pole of order 2 in one marked point $p_1=0$. This is equivalent to considering
    a diagonal generic meromorphic connection $d-\mathcal{A}$ on the trivial rank 2 vector bundle over
the unit disc $\mathbb{D} \subset \mathbb{C}$ with only a pole of order 2 at 0. For $u$ a coordinate vanishing at 0, we write
\begin{eqnarray}
\mathcal{A}= d Q +T_1 \frac{du}{u}\label{par1}
\end{eqnarray}
where $T_1$ is a constant diagonal matrix, and $Q$ is a diagonal matrix of meromorphic functions \cite[p.154]{BoalchIso}\footnote{Note that $Q$ is determined by $\mathcal{A}$ and $u$ by requiring that it
has constant term zero in its Laurent expansion with respect to $u$.}.
We write the complexified connection as:
\begin{eqnarray}
\mathcal{A}=  du \left(\frac{T_2}{u^2}+\frac{T_1}{u}\right). \label{L_wit1}
\end{eqnarray}
Hence, in terms of \cite{Witten:2007td}, one is taking $T_2=\frac{dQ}{du} u^2$.  Considering $\widetilde q_{ij}$ as in  \cite[Eq. (2.11)]{Witten:2007td} without the tilde, this matrix is
\begin{equation}
T_2= \left(\begin{array}{cc}\widetilde q_{12}&0\\0& \widetilde q_{22}\end{array}\right)\label{L_wit2}.
\end{equation}
One should note that along \cite{Witten:2007td}, when the order $k$ of the single pole is fixed, the notation is simplified and the diagonal entries of $T_k$ are denoted by $\widetilde q_j$ with no tilde.
For convenience
we shall follow the notation in \cite{BoalchIso} appearing in \eqref{par1}.    In particular, we have
\begin{eqnarray}
Q=\left(\begin{array}
{cc}
q_1(u)&0\\0&q_2(u)
\end{array}\right), \label{par2}
\end{eqnarray}
and define $q_{ij}(u)$ as the leading term of $q_i(u)-q_j(u)$. Letting $q_1(u) -q_2(u)=a/u+b$, then
 \begin{eqnarray}
 q_{12}(u)=a/u,~{\rm and }~
 q_{21}(u)=-a/u.\end{eqnarray}
 Note that from \eqref{L_wit1}-\eqref{L_wit2} one has that $\widetilde q_i = q'_i(u)u^2$, and hence $\widetilde q_i -\widetilde q_j = -a$.
 In what follows we shall  use polar coordinates and write
 $a=\widetilde r e^{i \widetilde \theta}.$

Associated to the above connections are \textit{anti-Stokes directions}, which can be obtained as follows. Consider directed lines in $\mathbb{C}$ through the origin parameterized by $S^1$. For ${\rm d}_1,{\rm d}_2\in S^1$ we let ${\rm Sect}({\rm d}_1, {\rm d}_2)$ be the (open) sector swept out by rays rotating   from the point ${\rm d}_1$ to the point ${\rm d}_2$.

\begin{definition}\label{anti}
The {\bf anti-Stokes directions} $\mathbb{A} \subset S^1$ obtained through the matrix $Q$ are the directions $d \in S^1$ for which either \begin{eqnarray}q_{12}(u) \in \mathbb{R}_{<0}, ~{~\rm or~}~
q_{21}(u) \in \mathbb{R}_{<0}
\end{eqnarray}
for $u$ on the ray specified by $d$.
\end{definition}

These are the directions along which $e^{(a/u + b)}$ (respectively $e^{-(a/u + b)}$) decays most rapidly as $u$ approaches 0 (note that $\mathbb{A}$ is independent of the coordinate choice $u$). Moreover, these directions  have $\pi$ rotational symmetry: if
$\pm ~\frac{a}{u} \in \mathbb{R}_{<0}$ then $\mp~\frac{a}{u} e^{-i\pi}\in \mathbb{R}_{<0}$, and thus $\# \mathbb{A}:=2.$
Note that if $q_{12}(u)\in  \mathbb{R}_{<0}$ for a direction $d$, then $q_{21}(u)\not \in  \mathbb{R}_{<0}$ for that direction (and vice versa), and thus in this set up,   a direction is referred to as a {\bf half-period}.
\begin{definition}\label{root}
For $d\in \mathbb{A}$ an anti-Stokes direction, the {\bf roots} of $d$ are
\[{\rm Roots}(d) :=\{(ij)~ | ~q_{ij}(u) \in  \mathbb{R}_{<0}~{\rm~ along~} d\}.\]
\end{definition}
 \begin{remark}
In order to find the directions, for one single pole of order $k$ we must study angles for which $a/u^{k-1}$ is negative. For $a=\widetilde r e^{i\widetilde \theta}$ as before, we need $u =re^{i\theta}$ for which $\widetilde \theta-(k-1)\theta=(2m+1)\pi $ for $m\in \mathbb{N}$. In such a case, one has that $a/u^{k-1}=-\widetilde r/r^{k-1}$ which is negative. Moreover, in general we shall call a {\bf half-period}  an $l$-tuple ${\rm d}=({\rm d}_1,\ldots, {\rm d}_l)\subset \mathbb{A}$ of consecutive anti-Stokes directions, where $2l=\# \mathbb{A}$. When weighted by their multiplicities, the number
of anti-Stokes directions in any half-period is $1={\rm Mult}({\rm d}_1)+\ldots+{\rm Mult}({\rm d}_l).$\end{remark}
In our case there are only two half periods (see Figure \ref{sec3}):
\begin{itemize}
\item One supported by $q_{12}(u)$ in a direction ${\rm d}_1:= e^{i(\widetilde \theta -\pi)},$ for which taking $u=r e^{i(\widetilde \theta-\pi)}$ one has $q_{12}(u)=-\frac{\widetilde r}{r}\in \mathbb{R}_{<0};$
\item One supported by $q_{21}(u)$ in a direction ${\rm d}_2:=e^{-i\pi} {\rm d}_1=e^{i\theta},$ for which taking $z=r e^{i\theta}$ one has $q_{21}(u)= -\frac{\widetilde r }{r}\in \mathbb{R}_{<0}.$
\end{itemize}

Moreover, from Definition \ref{root},   in the current setting one has
\begin{eqnarray}
{\rm Roots}({\rm d}_1) =\{(12)\},~{\rm ~and ~}~
{\rm Roots}({\rm d}_2) =\{(21)\}.
\end{eqnarray}
 \begin{definition}\label{factor}
The {\bf multiplicity} ${\rm Mult}(d)$ of $d$ is the number of roots supporting $d$, which from Definition \ref{root} is 1. Finally, the {\bf group of Stokes factors} \footnote{Note that these are called {\it Stokes matrices} in \cite{MartiniTime} and referred to as $St_d$.}
associated to $d$ is the group
\[\mathbb{S}to_d(\mathcal{A}):=\{K\in  G ~| ~(K)_{ij}=\delta_{ij} ~{~\rm unless~} ~(ij) ~{\rm~is~ a ~root ~of~} d\},\]
which is a unipotent subgroup of $G=GL(2,\mathbb{C})$ of dimension equal to  ${\rm Mult}(d)=1$.\end{definition}

\begin{figure}[h]
 \centering
  \includegraphics[width=0.3\textwidth]{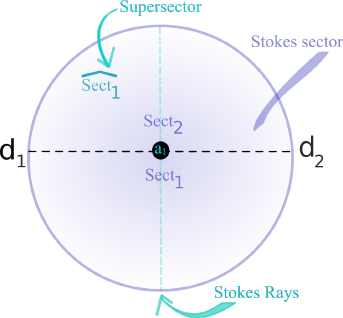}
\caption{Stokes rays, and Stokes Sectors and Super Sectors}\label{Figure 2}
\end{figure}

For both directions ${\rm d}_1$ and ${\rm d}_2$ that one has, the group of Stokes factors is a 1-dimensional subgroup given by
\begin{eqnarray}\mathbb{S}to_{{\rm d}_1}(\mathcal{A})=\left\{  \left(\begin{array}{cc}
1& \widetilde u\\
0&1
\end{array}\right)~{\rm for ~} \widetilde u \in \mathbb{C}\right\}~{\rm and ~} ~\mathbb{S}to_{{\rm d}_2}(\mathcal{A})=\left\{  \left(\begin{array}{cc}
1& 0\\
\widetilde u&1
\end{array}\right)~{\rm for ~} \widetilde u \in \mathbb{C}\right\}\end{eqnarray}
In particular, note that this agrees with Lemma 3.2 of \cite{BoalchIso} since, as a variety,
\begin{eqnarray}
\mathbb{S}to_{{\rm d}_1}(\mathcal{A})\cdot \mathbb{S}to_{{\rm d}_2}(\mathcal{A})&\cong& U_+ \times U_-\\
(K_1,K_2)&\mapsto&(S_1 , S_2 )\end{eqnarray}
for $U_\pm$ the lower and upper triangular unipotent subgroups of $GL(2,\mathbb{C})$, and
\begin{eqnarray}S_1:=K_{1}K_{2},~{\rm and}~ S_2:=K_{2}K_{1}.\end{eqnarray}
 We will see later that these are the {\bf Stokes matrices}, and are each in $U_{\pm}$ respectively (but one needs to be careful with the above double lower index).
Each of the two Stokes directions defines an ordering:
$q_1{<}_{{\rm d}_1}q_2$, and $q_2{<}_{{\rm d}_2}q_1.$
  In particular one has that
\begin{eqnarray}e^{q_1}/e^{q_2}&\rightarrow& 0~ {\rm ~along~ the~ ray~} \theta_1:=  e^{ i\frac{3\pi}{2}}\in  S^1~{\rm~ bisecting} ~{\rm Sect}_1:=Sect({\rm d}_1, {\rm d}_2),\nonumber \\
e^{q_2}/e^{q_1}&\rightarrow& 0~{\rm ~along~ the~ ray~} \theta_{2}:=e^{ i\frac{\pi}{2}}~ \in  S^1~~{~\rm~ bisecting} ~{\rm Sect}_2:=Sect({\rm d}_2, {\rm d}_1);\nonumber
\end{eqnarray}
and these rays are called {\bf Stokes Rays} as seen in Figure \ref{Figure 2}, and the sectors bounded by the {\bf Stokes rays} are referred to as   {\bf super sectors} $\widehat{{\rm Sect}}$.
\begin{remark}[Parameter space]
One should note that from \eqref{par1} the only parameters in all the above considerations are the constant $2\times 2$ diagonal matrix $T_1$, and the diagonal matrix $Q$. The latter, from \eqref{par2}, is equivalent to the parameters $a$ and $b$ for which $q_1(u)-q_2(z)=au+b$.
\end{remark}

We shall now consider the local moduli of meromorphic connections. For this, let ${\rm Sys}(\mathcal{A})$ be the set of germs at $0\in \mathbb{C}$ of meromorphic connections on the rank 2 bundle that are formally equivalent to $d-\mathcal{A}$ (as in \eqref{par1}).  Formally,
\[{\rm Syst}(\mathcal{A})=\{d-A ~| ~A=\widehat F[\mathcal{A}]~{\rm  for~ some}~\widehat F\in G[[z]]\},\]
 where $A$ is a matrix of germs of meromorphic one-forms, $G[[u]]=GL(2,\mathbb{C}[[u]])$ and \[\widehat F [\mathcal{A}]=(d\widehat F)\widehat F^{-1}+\widehat F\mathcal{A}\widehat F^{-1}.\]

For $G\{u\}:= GL(2, \mathbb{C}\{u\})$ the subgroup of germs of holomorphic bundle automorphisms, we are interested in
${\rm Sys}(\mathcal{A})/G\{u\}. $
For Higgs bundles of rank 1 or with just one pole,  this is just a point. Since we are considering here  rank 2 Higgs bundles with one single double pole, in this case the group is non-trivial and it can be described in terms of Stokes matrices. We shall denote by  $\widehat{{\rm Sys}}_{cf}(\mathcal{A}) $   the set of compatibly framed connection germs
with both irregular and formal type $\mathcal{A}$.

\begin{definition}
The set $\widehat{{\rm Sys}}(\mathcal{A})_{cf}$ is isomorphic to  the set of {\bf marked pairs}, defined as
\[\widehat{{\rm Sys}}(\mathcal{A})_{mp}:=\{(A,\widehat F)~|~A\in  {\rm Sys}(\mathcal{A}),~ \widehat F \in G[[u]], ~ A=\widehat F[\mathcal{A}]\}.\]
\end{definition}
 \begin{remark}\label{rem1}We can think of $\widehat{{\rm Sys}}(\mathcal{A})_{cf}$ as the set of pairs $(A,g_0)$ with $A\in {\rm Sys}(\mathcal{A})$ and $g_0\in G
$ such that $g_0[A]$ and $\mathcal{A}$ have the same leading term.
\end{remark}Note that  the above correspondence\footnote{The matrix entries $\widehat F_{ij}$ are 0 if $i\neq j$, and constant if $i=j$. } can be seen by taking $g_0:=\widehat F(0)^{-1}$. ~
Then, one has the quotient
$\mathcal{H}(\mathcal{A})=\widehat{{\rm Sys}}(\mathcal{A})/G\{u\},$
and for $T\cong (\mathbb{C}^*)^2$ it follows that
${\rm Sys}(\mathcal{A})/G\{u\}\cong \mathcal{H}(\mathcal{A})/ T.$
We shall  consider the following sectors ${\rm Sect}$ and super sectors $\widehat{{\rm Sect}}$: \begin{eqnarray}
{\rm Sect}_1&:=&{\rm Sect}(e^{i(\widetilde \theta -\pi)},e^{i\theta})\label{sec1}\\
{\rm Sect}_2&:=&{\rm Sect}(e^{i\theta},e^{i(\widetilde \theta -\pi)}) ~(~:={\rm Sect}_0)\label{sec2}\\
\widehat{{\rm Sect}}_1&:=&{\rm Sect}(e^{i(\widetilde \theta -\frac{3}{2}\pi)},e^{i(\widetilde\theta+\frac{1}{2}\pi)}) \ni {\rm d}_1\\
\widehat{{\rm Sect}}_2&:=&{\rm Sect}(e^{i(\widetilde\theta-\frac{1}{2}\pi)},e^{i(\widetilde \theta -\frac{1}{2}\pi)})\ni {\rm d}_2
\end{eqnarray}

Note that given $\widehat F$ a formal transformation such that
$A:=\widehat F[\mathcal{A}],$ there are canonical $2\times 2$ matrices of holomorphic functions
\begin{eqnarray}
\Sigma_1(\widehat F)~&{\rm on ~Sect}_1~{\rm such~that~}&~\Sigma_1(\widehat F)[\mathcal{A}]=A, \\
\Sigma_2(\widehat F)~&{\rm on ~Sect}_2~{\rm such~that~}&~\Sigma_2(\widehat F)[\mathcal{A}]=A,
\end{eqnarray}
defined uniquely  such that $\Sigma_j(\widehat F)$ can be analytically extended to $\widehat{\rm Sect}_j$, and such that it is asymptotic to $\widehat F$ on 0 within $\widehat{\rm Sect}_j$.
For a fixed $p\in {\rm Sect}_2$, choose    a branch of ${\rm log}(u)$ giving a lift  $\widetilde p$ of $p$ to the universal cover of the punctured disc $\mathbb{D}_0$. Then, the {\bf canonical fundamental solution} of $A$ on the sectors ${\rm Sect}_1$ and ${\rm Sect}_2$ are
 \begin{eqnarray}\Phi_1&=&\Sigma_1(\widehat F) u^{T_1}e^Q,\\
  \Phi_2&=&\Sigma_2(\widehat F) u^{T_1}e^Q,
 \end{eqnarray}
 respectively, for a fixed\footnote{Recall that from \eqref{par1} fixing $\mathcal{A}$ is equivalent to fix $Q$ and $T_1$. To this data that was already fixed, one is adding the data of Remark \ref{rem1} and a point $p$ with its lift $\widetilde p$, the latter being equivalent to the point $b_1$ in Figure  \ref{points}.} $(\mathcal{A},z,\widetilde p)$ and $(A,g_0)\in \widehat{Syst}(\mathcal{A})$ (as in Remark \ref{rem1}). For this fixed data, the {\bf Stokes factors} are
 \begin{eqnarray}
 K_1&=&e^{-Q}u^{-T_1}\kappa_1u^{T_1} e^{Q}, \\
 K_2&=&e^{-Q}u^{-T_1}\kappa_2u^{T_1} e^{Q},
 \end{eqnarray}
 where the $\kappa_1,\kappa_2$ are the matrix of holomorphic functions  defined by
  \begin{eqnarray}
 \kappa_1=  \Sigma_1(\widehat F)^{-1}\circ \Sigma_2(\widehat F),~{\rm ~and~}~
 \kappa_2= \Sigma_2(\widehat F)^{-1} \circ \Sigma_1(\widehat F).
 \end{eqnarray}

   The {\bf Stokes matrices} are essentially the transition matrices between the canonical fundamental solution $\Phi_2$ on ${\rm {\rm Sect}_0}$ and $\Phi_1$ on the opposite sector, ${\rm Sect}_1$, when they are continued along the two possible paths in the punctured disk joining these sectors. Moreover,    \begin{eqnarray}
  \Phi_1=\Phi_2 \cdot PS_-P^{-1}~{\rm ~and~}~
    \Phi_2=\Phi_1\cdot PS_+P^{-1} M_0,
  \end{eqnarray}
where $P_{ab}=\delta_{c(a)b}$ for $c$ the permutation of $\{1,2\}$ giving the ordering of the sector taken. (so it is the identity matrix if we consider the sector giving the ordering $q_1<q_2$).

\begin{remark}
Whilst understanding tentacles will be most useful when many poles are considered at the same time, we should note that if one needed to define a tentacle involving poles of order two, the extra data needed would be two points ${\rm b}_1$ and ${\rm b}_2$ in red   in Figure \ref{tent1}.

\end{remark}

\subsection*{Rank 2 Higgs bundles with 1 pole of order 4}\label{one4}

\begin{figure}[h]
 \centering
  \includegraphics[width=0.35\textwidth]{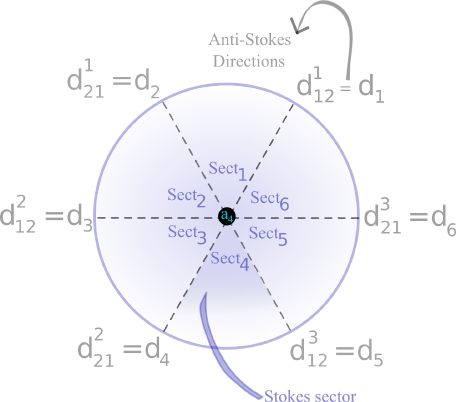}
\caption{Anti-Stokes directions.}\label{sec3}
\end{figure}

We shall focus here on Higgs bundles $(E,\Phi)$ on $\PP$ with one pole of order 4. Consider   a diagonal generic meromorphic connection $d-\mathcal{A}$ on the trivial rank 2 vector bundle over
the unit disc $\mathbb{D} \subset \mathbb{C}$ with a pole of order 4 at 0 and no others. For $u$ a coordinate vanishing at 0, as in \eqref{par1} we write
\begin{eqnarray}
\mathcal{A}= d Q +T_1 \frac{du}{u}
\end{eqnarray}
where $T_1$ is a constant diagonal matrix, and $Q$ is a diagonal matrix of meromorphic functions. In terms of the notation of Section 4 In particular, we have
\begin{eqnarray}
Q=\left(\begin{array}
{cc}
q_1(u)&0\\0&q_2(u)
\end{array}\right)
\end{eqnarray}
and define $q_{ij}(u)$ as the leading term of $q_i(u)-q_j(u)$. Letting $$q_1(u) -q_2(u)=\frac{a}{u^3}+\frac{b}{u^2}+\frac{c}{u}+d,$$ then  $q_{12}(u)=\frac{a}{u^3}$ and $q_{21}(u)=-\frac{a}{u^3}.$
We shall use polar coordinates and consider
 $a=\widetilde r e^{i \widetilde \theta}$.
 As before, the {\bf anti-Stokes directions} $\mathbb{A} \subset S^1$ are the directions $d \in S^1$ for which either \begin{eqnarray}q_{12}(u) \in \mathbb{R}_{<0}, ~{~\rm or~}~
q_{21}(u) \in \mathbb{R}_{<0}
\end{eqnarray}
for $u$ on the ray specified by $d$. These have $\pi/3$ rotational symmetry, since if $q_{ij}(u)\in \mathbb{R}_{<0}
$
then $q_{ji}(u ~e^\frac{\pi i}{3}))\in \mathbb{R}_{<0}$. In particular, this says that the number $r$ of anti-Stokes directions is divisible by 6, and so we define $l=r/6$. An $l$-tuple ${\bf d}=({\rm d}_1,{\rm d}_2,\ldots,{\rm d}_l)$ of consecutive anti-Stokes directions is a half period, and since
\[1={\rm Mult}({\rm d}_1)+\ldots+{\rm Mult}({\rm d}_l)\]
and multiplicities (number of roots supporting the direction $d$) are positive integers, one has that $l=1$ and hence there are $6$ anti-Stokes directions.
 Each of these directions form what is called {\it half-period}. The 6 anti-Stokes directions can be described as follows:

 \begin{itemize}
 \item Three directions are supported by $q_{12}(u)=  \widetilde r e^{i \widetilde \theta}/u^3$. For $j=1,2,3$, the
  directions are
  $ {\rm d}^j_{12}:= e^{i\left(\frac{\widetilde \theta}{3} - \frac{(2j+1)\pi}{3} \right)},$
  for which taking $u=r e^{i\left(\frac{\widetilde \theta}{3} -(2j+1) \frac{\pi}{3} \right)}$ one has $$q_{12}(u)=-\frac{\widetilde r}{r^3}\in \mathbb{R}_{<0};$$

  \item Three directions are supported by $q_{21}(u)=  -\widetilde r e^{i \widetilde \theta}/u^3$, defined as
  $ {\rm d}^j_{21}:= e^{\frac{i\pi}{3}} {\rm d}^j_{12}$ or equivalently ${\rm d}^j_{21}=e^{i\left(\frac{\widetilde \theta}{3} - \frac{2j\pi}{3} \right)},$
  for which taking $u=e^{i\left(\frac{\widetilde \theta}{3} - \frac{2j\pi}{3} \right)}$ one has $$q_{21}(u)=- \frac{\widetilde r}{r^3}\in \mathbb{R}_{<0};$$
 \end{itemize}
    and they appear in Figure \ref{sec3}, agreeing with Definition \ref{anti}.
%
 Recall from Definition \ref{factor} that the {\bf group of Stokes factors}
associated to a direction $d$ is the group
\[\mathbb{S}to_d(\mathcal{A}):=\{K\in  G ~| ~(K)_{ij}=\delta_{ij} ~{~\rm unless~} ~(ij) ~{\rm~is~ a ~root ~of~} d\},\]
  which is a unipotent subgroup of $G=GL(2,\mathbb{C})$ of dimension 1.
For all directions $ {\rm d}^j_{12}$ and $ {\rm d}^j_{21}$ that one has in this setting, for $j=1,2,3$, the group of Stokes factors is a 1-dimensional subgroup given by
\begin{eqnarray}\mathbb{S}to_{ {\rm d}^j_{12}}(\mathcal{A})=\left\{  \left(\begin{array}{cc}
1& \widetilde z\\
0&1
\end{array}\right)~{\rm for ~} \widetilde z \in \mathbb{C}\right\}\end{eqnarray}
\begin{eqnarray}\mathbb{S}to_{ {\rm d}^j_{21}}(\mathcal{A})=\left\{  \left(\begin{array}{cc}
1& 0\\
\widetilde u&1
\end{array}\right)~{\rm for ~} \widetilde u \in \mathbb{C}\right\}\end{eqnarray}
 In particular, this agrees with Lemma 3.2 (2) of \cite{BoalchIso} since, as a variety,
\begin{eqnarray}
\Pi_{j=1}^3\left( \mathbb{S}to_{ {\rm d}^j_{12}}(\mathcal{A})\cdot \mathbb{S}to_{ {\rm d}^j_{21}}(\mathcal{A})\right)&\cong& (U_+ \times U_-)^3\\
(K_1,\ldots, K_6)&\mapsto&(S_1, \dots, S_6) \end{eqnarray}
for $U_\pm$ the lower and upper triangular unipotent subgroups of $GL(2,\mathbb{C})$, and
\begin{eqnarray} S_i&:=&K_{i}K_{2(i-1)}\in U_{+/-}~{\rm for} ~i~{\rm ~odd/even}\label{Smat}\end{eqnarray}
the {\bf Stokes matrices}.
Note that, independently of $j$,  as in the previous case the Stokes directions $ {\rm d}^{j}_{12}$ and $ {\rm d}^{j}_{21}$ define the orderings
$q_1<_{ {\rm d}^{j}_{12}} q_2,$ and
$q_2<_{ {\rm d}^{j}_{21}}q_1.
$
Recall that for a rank 2 Higgs bundle with one pole of order 2 one had two sectors defined as in \eqref{sec1}-\eqref{sec2}. In this case, with only one pole of order 4 there are 6 sectors,  ${\rm {\rm Sect}_i} :={\rm Sect}({\rm d}_i , {\rm d}_{i+1} )
$, with indices mod 6 as in Figure \ref{sec3}.

In order to consider the local moduli of meromorphic connections, let ${\rm Sys}(\mathcal{A})$ be the set of germs at $0\in \mathbb{C}$ of meromorphic connections on the rank 2 bundle that are formally equivalent to $d-\mathcal{A}$.  Formally,
\[{\rm Syst}(\mathcal{A})=\{d-A ~| ~A=\widehat F[\mathcal{A}]~{\rm  for~ some}~\widehat F\in G[[u]]\},\]
 where $A$ is a matrix of germs of meromorphic one-forms, $G[[u]]=GL(2,\mathbb{C}[[u]])$ and \[\widehat F [\mathcal{A}]=(d\widehat F)\widehat F^{-1}+\widehat F\mathcal{A}\widehat F^{-1}.\]
As in the previous cases, one has  $\widehat{{\rm Sys}}_{cf}(\mathcal{A}) $   the set of compatibly framed connection germs
with both irregular and formal type $\mathcal{A}$.
The set $\widehat{{\rm Sys}}(\mathcal{A})_{cf}$ is isomorphic to  the set of {\bf marked pairs} be
\[\widehat{{\rm Sys}}(\mathcal{A})_{mp}:=\{(A,\widehat F)~|~A\in  {\rm Sys}(\mathcal{A}),~ \widehat F \in G[[u]], ~ A=\widehat F[\mathcal{A}]\}\]
 Consider $\widehat F$ a formal transformation such that
$A:=\widehat F[\mathcal{A}]$ has convergent series. From \cite[Theorem 3.1]{BoalchIso} there are canonical $2\times 2$ matrices of holomorphic functions on each   ${\rm Sect}_j$ given by
$
\Sigma_j(\widehat F)~{\rm on ~Sect}_i~{\rm such~that~}~\Sigma_j(\widehat F)[\mathcal{A}]=A,$
defined uniquely  such that $\Sigma_j(\widehat F)$ can be analytically extended to $\widehat{\rm Sect}_j$, and such that it is asymptotic to $\widehat F$ on 0 within $\widehat{\rm Sect}_j$.

\begin{remark}
As explained in \cite{BoalchIso}, of all the holomorphic isomorphisms between $\mathcal{A}$ and $A$ which are asymptotic to $\widehat F$, from the above analysis one is being chosen in a canonical way.
\end{remark}

  The {\rm canonical fundamental solution} of $A$ on the sectors ${\rm Sect}_j$ are
$\Phi_j:=\Sigma_j(\widehat F) u^{T_1}e^Q, $
 for a fixed $(\mathcal{A},u,\widetilde p)$ and $(A,g_0)\in \widehat{Syst}(\mathcal{A})$, where for a fixed $p\in {\rm Sect}_0$ one has chosen a branch of ${\rm log}(u)$ giving a lift  $\widetilde p$ of $p$ to the universal cover of the punctured disc $\mathbb{D}_0$. For this fixed data the {\bf Stokes factors} are constant invertible matrices\footnote{Since $u^{T_1} e^{Q}$ is a fundamental solution of $\mathcal{A}$ (i.e. its columns are a basis of solutions) we have $d(K_i)=0$.}
 \begin{eqnarray}
 K_j&=&e^{-Q}u^{-T_1}\kappa_ju^{T_1} e^{Q} \in \mathbb{S}to_{{\rm d}_j}(\mathcal{A}),
  \end{eqnarray}
 where the $\kappa_j$ are matrix of holomorphic functions  defined by
  \begin{eqnarray}
 \kappa_j&=&  \Sigma_j(\widehat F)^{-1}\circ \Sigma_{j-1}(\widehat F)
  \end{eqnarray}
\begin{remark}
Note that $\kappa_j[\mathcal{A}]=\mathcal{A}$, and thus it is an automorphism of $\mathcal{A}$. Moreover, $K_6 \cdots K_1= P S_6\cdots S_1P^{-1},$
where $P$ is the identity matrix when the ordering is   $q_1<q_2$.\end{remark}
If $\Phi_j$ is continued across the anti-Stokes ray ${\rm d}_{j+1}$, then on the sector  ${\rm Sect}_{j+1}$ it follows  that  $K_{j+1} :=\Phi^{-1}_{j+1} \circ \Phi_j$ for all $j$ except $K_1 :=\Phi^{-1}_{j+1} \circ \Phi_j\circ M_{0}^{-1}$ for $i=6$,
where as in Definition \ref{mono11} one has that  $M_0 := e^{2\pi i T_1}$  is the so-called {\it formal monodromy}.

\begin{figure}[h]
 \centering
  \includegraphics[width=0.4\textwidth]{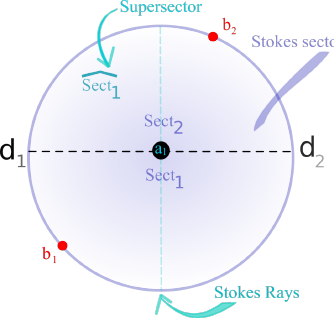}
\caption{Data needed to define a tentacle.}\label{points}
\end{figure}

   The {\bf Stokes matrices} $S_j=K_{j}K_{2(j-1)}$
   are essentially the transition matrices between the canonical fundamental solutions: if $\Phi_j$ on ${\rm {\rm Sect}_j}$ is continued onto   ${\rm Sect}_{j+1}$,  then  for $j=1,\ldots, 5$ one has $
  \Phi_j=\Phi_{j+1}  \cdot PS_{j+1}P^{-1}
$ where $P_{ab}=\delta_{c(a)b}$ for $c$ the permutation giving the ordering of the sector taken (so it's the identity matrix if we stick with the sector giving the ordering $q_1<q_2$).
For $j=6$ one has that
$  \Phi_6=\Phi_{1} \cdot PS_{1}P^{-1}M_0.
$\begin{remark}
If one continues $\Phi_0=\Phi_6$ around the sectors and back to ${\rm Sect}_0$, it will become
\begin{eqnarray}
 \Phi_{0}  \cdot PS_{6}\cdots S_2 S_1 P^{-1}M_0
   \end{eqnarray}
\end{remark}
If a wild Higgs bundle had multiple poles, one of which of order 4, then the extra data needed to define a tentacle coming from the order 4 pole would be as   in red in Figure \ref{points}.

\subsection*{Rank 2 Higgs bundles with 2 poles of order 2}\label{two2} When considering Higgs bundles with multiple poles of different orders, one needs to have a coherent way of encoding the data coming from each pole. Following \cite{BoalchIso}, we shall recall here how this is done through the so-called tentacles in the case of a rank 2 Higgs bundles with exactly two poles of order 2.

In a more general setting, consider Higgs bundles with different poles over a divisor $D=\{p_1,\ldots,p_4\}$ consisting of at most four points on $\mathbb{P}^1$, and let $V$ be a homomorphically trivial bundle over $\mathbb{P}^1$. Then, we may consider $u$ a local coordinate which vanishes at $p_i$, and near each point $p_i$ there is a local trivialization of $V$. Then, near each $p_i$ let $\nabla=d-\mathcal{A}$ be a meromorphic connection with poles on each $p_i\in D$ of order $k_i$: one may write
\begin{eqnarray}
\mathcal{A}=\sum_{i=1}^{4} dQ_i+T^i_1\frac{du}{(u-p_i)}+{holomorphic ~terms},\end{eqnarray}
or equivalently, following the notation of Section 4, as
\begin{eqnarray}
\mathcal{A}=\sum_{i=1}^{4}\left(  \frac{T^i_{k_i}}{(u-p_i)^{k_i}}+\cdots +\frac{T^i_{1}}{(u-p_i)^{1}}\right),
\end{eqnarray}
where $T^i_j$ are $2\times 2$ matrices of $(1,0)$ forms.
As mentioned before,  the connection $\nabla$ has a pole of order $k_i$ at $p_i$, and assuming that $T^1_1+\cdots+^4_1=0$, it has no other poles.   To deal with this type of Higgs bundles  we choose disjoint open discs ${\rm d}_i$ on $\mathbb{P}^1$ with $p_i\in {\rm d}_i$ and, for each $i$, a coordinate $u_i$ on ${\rm d}_i$ vanishing at $p_i$. Thus the local picture above of the previous sections is repeated on each such disc.

\begin{remark}
The Stokes data associated to each pole and its disc is put together through {\it tentacles} (see \cite{BoalchIso} where these were defined and thoroughly studied), and in what follows we shall review how these are obtained. In particular, one needs to fix a base point $p_0\neq p_i$, and a point in each sector.
\end{remark}

\begin{figure}[h]
 \centering
  \includegraphics[width=0.55\textwidth]{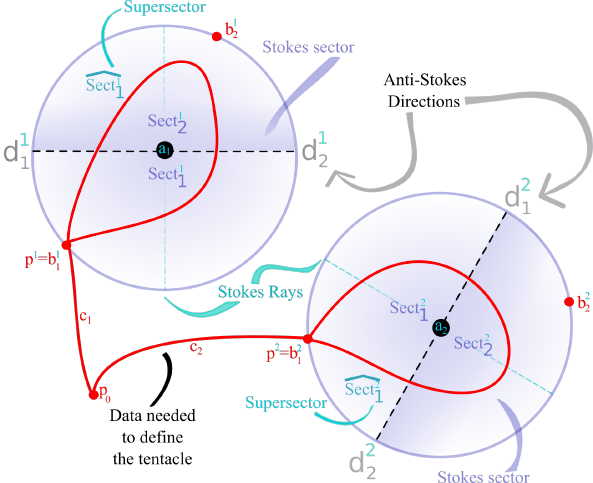}
\caption{Tentacles for two poles of order 2.}\label{tent1}
\end{figure}

Consider then a fixed base point $p_0\neq p_i$,  and a choice $b_{\xi}$ in each sector bounded by anti-stokes directions at each point $p_i$. These points will allow one to define the tentacle associated to the Higgs bundle, which ultimately will consist of the data   in Figure \ref{tent1}.

Define \textbf{\textit{monodromy manifolds}}   as spaces of representations of the  groupoid $\widetilde\Gamma$:
\begin{itemize}
\item The objects of $\widetilde\Gamma$ are the elements of $\widetilde B :=\{p_0\}\cup \widetilde B_ 1 \cup \widetilde B_ 2\cup \widetilde B_ 3\cup \widetilde B_ 4$, where  $\widetilde B_i$ is the discrete set of points of the   universal cover of the punctured disc  ${\rm d}_i/p_i$ which are above one of the $b_{\xi}$'s.
\item   If $\widetilde p_1, \widetilde p_2\in \widetilde B$, then the set of morphisms of $\widetilde \Gamma$ from $p_1$ to $p_2$ is the set of homotopy classes of paths $\gamma:[0,1]\rightarrow \mathbb{P}^1-\{p_1,p_2,p_3,p_4\}$ from $p_1$ to $p_2$. Here we have denoted by $p\in \mathbb{P}^1$ the underlying point of $\widetilde p \in \widetilde B$ (namely $p_0$ or one of the $b_{\xi}$).
\end{itemize}
For a path $\gamma_{p_2p_1}$ from  $p_1$ to $p_2$, we shall denote by $[\gamma_{p_2p_1}]\in \widetilde \Gamma$ its class. Each choice of basis of the fibre $V_{p_0}$ of the rank 2 bundle $V$ at $p_0$, one has that $(V, \nabla)$ naturally determines a representation of the groupoid $\Gamma$ in the group $G=GL (2,\mathbb{C})$, as follows.
Since there is a canonical choice of basis $\Phi:\mathbb{C}^2\rightarrow V$ of $\nabla$-horizontal sections of $V$ in a neighbourhood of $p_{i}$, for $i=1,2$, one can extend these bases along the path $\gamma_{p_2p_1}([0,1])$. Since they are both horizontal, one has that $\Phi_1=\Phi_2 \cdot C$ on this path, for some $C\in GL (2,\mathbb{C})$. Then, the representation is
$\rho([\gamma_{p_2p_1}]):=C=\Phi_2^{-1}\Phi_1.$
In particular, $\rho$ encodes all possible {\it connection matrices} between sectors at different poles as well as all the {\it Stokes factors} and {\it Stokes matrices} at each pole. Henceforth, we shall denote by $\gamma_i$ a path connecting $p_0$ to $p_i$, as appearing in red in Figure \ref{tent1}.
The representations have two properties to which we shall allude in the coming sections (see \cite[Lemma 3.6]{BoalchIso})
\begin{itemize}
\item[{\bf (SR1)}]For any $i$, if $p_1\in \widetilde B_i$ and $p_2$ is the next element of $B_i$ after $p_1$ in a positive sense, consider  $\gamma_{p_1p_2}$  a small arc in ${\rm d}_i$ from $p_1$ to $p_2$. Then  $\rho(\gamma_{p_2p_1})\in \mathbb{S}to_d(A_i^0)$, where $d$ is the unique anti-Stokes ray that $\gamma_{p_2p_1}$  crosses.

\item[{\bf (SR2)}]For each $j$ there is a diagonal matrix $\Lambda_j$ (with distinct eigenvalues mod $\mathbb{Z}$ if $k_j =1$) such that for any $p_1\in \widetilde B_j$, $p_2\in \widetilde B$ and morphism $\gamma_{p_2p_1}$, one has that
$\rho(\gamma_{p_2(p_1+2\pi)})=\rho(\gamma_{p_2p_1})\cdot e^{2\pi i ~^j\Lambda},
$ where $\gamma_{p_2(p_1+2\pi)}=\gamma_{p_2p_1}$ are equal as paths, but $(\widetilde p_1+2\pi)$ is the next point of $\widetilde B_j$ after $\widetilde p_1$ (in a positive sense) which is also above $p_1$.
\end{itemize}

 \begin{definition}
 A {\bf Stokes representation} $\rho$ is a representation of the groupoid $\widetilde \Gamma$ into $GL(2,\mathbb{C})$ together with a choice of 4 diagonal matrices $^j\Lambda$ such that {\bf (SR1)} and {\bf (SR2)} hold. The set of Stokes representations will be denoted $Hom_{\mathbb{S}} (\widetilde \Gamma , GL(2,\mathbb{C}))$.
 \end{definition}
\begin{definition}
The matrices $^j\Lambda$ associated to a Stokes representation $\rho$ will be called the {\bf exponents of the formal monodromy} of $\rho$ and the number
$\deg(\rho):= \sum_{j}{\rm Tr}(^j\Lambda )$ is the degree of $\rho$.
\end{definition}
In order to define the Stokes matrices in the case of multiples poles (through a definition consistent with \eqref{Smat}), one needs to make choices of points and paths between them. These choices are encoded in a choice of the so-called \textit{\textbf{tentacle}} appearing in Figure \ref{tent1}:
\begin{enumerate}
\item A point $p_j$ in some sector at $p_j$ between two anti-Stokes rays,
 for $j=1, \ldots, 4$.
\item A lift $\widetilde p_j$ of each $p_j$ to the universal cover of the punctured disc
${\rm d}_j /\{p_i\}$.
\item A base-point $p_0 \in \mathbb{P}^1/\{p_1,\ldots, p_4\}$.
\item A path $\gamma_j: [0, 1] \rightarrow \mathbb{P}^1/\{p_1,\ldots, p_4\}$ in the punctured sphere, from $p_0$ to $p_j$ for each  $j$, such that the loop
$(\gamma^{-1}_4\beta_4\gamma_4)(\gamma^{-1}_3\beta_3\gamma_3)(\gamma^{-1}_2\beta_2\gamma_2)(\gamma^{-1}_1\beta_1\gamma_1)
$ based at $p_0$ is contractible in $\mathbb{P}^1/\{p_1,\ldots, p_4\}$, where $\beta_j$ is any loop in ${\rm d}_j/\{p_j\}$ based at $p_j$ encircling $p_j$ once in a positive sense.
\end{enumerate}

\begin{remark}
Note that a choice of tentacle implies a labelling $P_j$ of the permutation associated to $p_j$ (which for simple poles is the identity matrix).
\end{remark}
 In the case of only one pole, one can still define the tentacle data associated to the system, which is consistent with what was described in previous sections.
 We shall denote by $C_j:=P^{-1}_j\rho(\gamma_j)\in GL(2,\mathbb{C})$, for $j=1,2,3,4$, where as before $\gamma_j$ is the path connecting $p_0$ to $p_j$. Moreover, denote by $^i\sigma_j$ the   morphism from $^i \widetilde b_{(j-1)\cdot l}$ to $^i \widetilde b_{j\cdot l}$ with
underlying path a simple arc in ${\rm d}_i /\{p_i\}$ from $^i  b_{(j-1)\cdot l}$ to $^i  b_{j\cdot l}$  in a positive
sense (where $l=l_i=r_i/(2k_i -2)$).

One  defines the {\bf Stokes matrices} by the formulae (which agrees with the case of just one pole in \eqref{Smat}):
\begin{eqnarray}
 ^iS_1:=P^{-1}_{i}\rho(^i\sigma_0)~^iM_{0}^{-1}P_i.
~{\rm ~and~}~^iS_j:=P^{-1}_{i}\rho(^i\sigma_j)P_i~{\rm for} ~j=2,...,2k_i-2
 \end{eqnarray}

In the case of a rank 2 Higgs bundles with two poles of order two as we are considering in this section, there are only two marked points $p_1$ and $p_2$ and
  \begin{eqnarray}
\nabla:=d- \left( T^1_{2} \frac{du}{(u-p_1)^{2}}  +T^1_{1} \frac{du}{(u-p_1)}\right)- \left( T^2_{2} \frac{du}{(u-p_2)^{2} } +T^2_{1} \frac{du}{(u-p_2)}\right).
\end{eqnarray}
In this case one considers two discs ${\rm d}_i$ containing $p_i$, for $i=1,2$ and choose $p_0\in \mathbb{P}^1-\{p_1,p_2\}$ and $b^i_j$, for $i,j=1,2$, marked points in each of the two sectors that each disc ${\rm d}_i$ has.   Then, morphisms of $\widetilde \Gamma$ correspond to paths connecting two points in $\{b^i_j, p_0\}_{i,j=1,2}$ not passing through $p_1,p_2$, and the tentacle is given by the data appearing in Figure \ref{tent1}.

\begin{remark} One should note that in this case the tentacle is determined by the choice of the discs ${\rm d}_1$, ${\rm d}_2$, together with a base point $p_0$, two points ${\rm p}^i={\rm b}^i_1$ and paths ${\rm c}_i$ from ${\rm p}_0$ to ${\rm p}_i$. Taking the canonical orientation the matrices $P_i$ yields the $2 \times 2$ identity matrices.
\end{remark}

\section{An Integrable System} \label{app:iscft}

In this section, we collect a few results from \cite{Donagi:1995am}
which will make the argument leading to (\ref{Mtot}) more precise.  The argument is a straightforward extension
of an argument in \cite{Witten:2007td}, which also referred to \cite{Donagi:1995am}, but not explicitly. We primarily work
in terms of a gauge group $U(N)$ and its complexification
$GL(N,\mathbb{C})$ so we state those first, then adapt them to $SL(N,\mathbb{C})$.

We consider Higgs bundles on $C$ valued in $L=K_C(D)$, where $D$ is an effective divisor.  So
we have a rank $N$ and degree $d$ vector bundle $E$ on $C$ and  a Higgs bundle structure
\begin{equation}
\Phi:E\to E\otimes K_C(D).
\end{equation}
If $D=\sum m_ip_i$, then $\Phi$ can be identified with a Higgs field having poles of order at most $m_i$ at $p_i$.
Let
\begin{equation}
B_L=H^0(C,K_C)\oplus H^0(C,K_C^2)\oplus\ldots H^0(C,K_C^r).
\end{equation}
The characteristic polynomial of $\Phi$ gives rise to a
map $\mathrm{Higgs}_C(N,d,L)\to B_L$, where $\mathrm{Higgs}_C(N,d,L)$ is the moduli space of Higgs bundles of the above type. We can define spectral covers for $L$-valued Higgs fields in perfect analogy to the case $D=0$ of the Hitchin  system.

\bigskip\noindent
{\bf Theorem.} Suppose that $L^{\otimes N}$ is very ample and $\deg(D)>\mathrm{max}(2,\rho)$,
where $0\le\rho<r$ is the residue of $d$ mod $N$.  Then
\begin{enumerate}
\item
The moduli space $M_C(N, d, L)$ of stable $L$-
valued Higgs fields of rank $N$ and degree $d$ has a smooth component
$M_C^{\mathrm{sm}}(N, d, L)$ of top dimension
$N^2(2g -2 + deg(D)) + 1 + \epsilon_D$, where $\epsilon_D$ is 1 if $D = 0$ and zero if
$D > 0$.  Furthermore, $M_C^{\mathrm{sm}}(N, d, L)$ is the unique component  containing
those Higgs
pairs which are supported on irreducible and reduced spectral curves.
\item
$M_C^{\mathrm{sm}}(N, d, L)$ has a canonical Poisson structure (depending on $D$).
\item  The characteristic polynomial map
$H : M_C^{\mathrm{sm}}(N, d, L) \to B_L$
is an algebraically
completely integrable Hamiltonian system. The generic (Lagrangian)
fiber is a complete Jacobian of a smooth spectral curve of genus
$N^2(g-1) + 1 +
(deg D)N(N-1)/2$.
\end{enumerate}
All dimensions in the theorem are complex. Observe that
the dimension of the fiber is computed readily by Riemann-Hurwitz for the spectral cover.
The dimension of the base is computed by Riemann-Roch, and then the dimension of the total
space is found by adding the dimensions of the base and fiber.

Similar results appear in \cite{Bottacin}, which also contains results for bundles with
fixed determinant, i.e.\ $SL(N,\mathbb{C})$-bundles.  In this case, the base is replaced by
\begin{equation}
B_L=H^0(C,K_C^2)\oplus H^0(C,K_C^3)\oplus\ldots H^0(C,K_C^N),
\end{equation}
and the fiber is replaced by a torsor for the Prym variety.  Since we are interested in
poles, we assume that $D\ne0$.  The dimension of the base is
computed to be
\begin{equation} \label{eq:basedim}
(N^2-1)(g-1)+\frac{(N+2)(N-1)}2\deg(D).
\end{equation}
By subtracting $g$ from the dimension of the Jacobian, the dimension of the Prym is 
\begin{equation}\label{eq:fiberdim}
(N^2-1)(g-1)+\frac{(N(N-1)}2\deg(D),
\end{equation}
from which the dimension of the component of the moduli space is
\begin{equation}\label{eq:totaldim}
(N^2-1)(2g-2+\dim(D)).
\end{equation}
Specializing to $\mathbb{P}^1$, these can be summarized as the following theorem, where we wrote $M_{\mathbb{P}^1,0}(N,L)$ for the
moduli space of $SL(N,\mathbb{C})$ Higgs bundles on $\mathbb{P}^1$ valued in $L$.  Since $d=0$, the
only condition is that $L$ is very ample, or equivalently $\deg(D)\ge3$.

\bigskip\noindent
{\bf Theorem.} Suppose that $\deg(D)\ge3$.  Then
\begin{enumerate}
\item
The moduli space $M_{\mathbb{P}^1,0}(N, L)$ of stable $L$-
valued $SL(N,\mathbb{C})$ Higgs bundles has a smooth component
$M_{\mathbb{P}^1,0}^{\mathrm{sm}}(N, L)$ of top dimension
$(N^2-1)(2g-2+\dim(D))$.
\item  The characteristic polynomial map
$H : M_C^{\mathrm{sm}}(N, d, L) \to B_L$
is an algebraically
completely integrable Hamiltonian system. The generic (Lagrangian)
fiber is a torsor for the Prym variety of a smooth spectral cover,
an abelian variety of dimension
$(N^2-1)(g-1)+\frac{(N(N-1)}2\deg(D)$.
\end{enumerate}

\section{Complex Structure Deformations} \label{app:cplx}

In this Appendix, we describe complex structure deformations, extending the analysis in \cite{Anderson:2013rka}. We implicitly make reference
to these features when we translate background values for hypermultiplets to complex structure deformations.

We begin by recalling the relevant tools of deformation theory of a possibly
singular algebraic variety $X$.  Let $\Omega^1_X$ be the sheaf of holomorphic 1-forms
on $X$.
Then we put
\begin{equation}
\mathcal{T}^0_X=\underline{Hom}_X(\Omega^1_X,\mathcal{O}_X),
\qquad
\mathcal{T}^1_X=\underline{Ext}^1_X(\Omega^1_X,\mathcal{O}_X).
\end{equation}
Then $\mathcal{T}^0_X$ is just the tangent sheaf of $X$, which is a vector bundle if
$X$ is smooth but not in general.
Furthermore, if $X$ is smooth, then $\mathcal{T}^1_X$ is zero because $\Omega^1_X$ is
a vector bundle.  Since all Ext sheaves
are defined locally on $X$, we see that $\mathcal{T}^1_X$
is supported on the singular locus $\mathrm{Sing}(X)$ of $X$.

The first order deformations of $X$ are given by $\mathrm{Ext}^1_X(\Omega^1,\mathcal{O}_X)$
(the Ext group, not the Ext sheaf).  The local to global spectral sequence for
$H^p(X,\underline{Ext}^q_X(\Omega^1_X,\mathcal{O}_X))\implies Ext^q_X
(\Omega^1_X,\mathcal{O}_X))$
gives
\begin{equation}
0\to H^1(X,\mathcal{T}^0_X)\to \mathrm{Ext}^1_X(\Omega^1_X,\mathcal{O}_X)
\to H^0(X,\mathcal{T}^1_X)\stackrel{\delta}{\to} H^2(X,\mathcal{T}^0_X).
\label{eq:localtoglobal}
\end{equation}
The map $\mathrm{Ext}^1_X(\Omega^1,\mathcal{O}_X)
\to H^0(X,\mathcal{T}^1_X)$ takes a global first order deformation of $X$ to
the associated local deformation of a neighborhood of the singularity.
Thus $H^1(X,\mathcal{T}^0_X)$ is the space of first order deformations of $X$
which preserve $\mathrm{Sing}(X)$, and furthermore, (\ref{eq:localtoglobal})
says that a first order deformation $\rho\in H^0(X,\mathcal{T}^1_X)$ of a
neighborhood of $\mathrm{Sing}(X)$ extends to a first order deformation
of all of $X$ if and only if $\delta(\rho)=0$.

For emphasis, the domain $H^0(X,\mathcal{T}^1)$ of the map $\delta$
depends only on a neighborhood
of $\mathrm{Sing}(X)$, but since $\mathcal{T}^0_X$ is supported on all of $X$,
$\delta$ depends on the global
geometry of $X$, not just a neighborhood of $\mathrm{Sing}(X)$.
We point out that this is a familiar situation in string
theory and in geometry, appearing in a compact conifold transition
\cite{clemens,friedman,Greene:1995hu}.

Now suppose we have a curve $C$ of $A_1$ singularities
in a Calabi-Yau threefold $X$, so that $\mathrm{Sing}(X)=C$.
Since $\mathcal{T}^1_X$ is local, we pick a local model for $X$ to be
given by the equation $xy=z^2$ in the total space $Y$ of the bundle $L_1\oplus L_2\oplus K_C$,
where $L_1\otimes L_2\simeq K_C^2$.
Letting $\mathcal{I}_X\subset \mathcal{O}_Y$ be the ideal sheaf
of $X$, we have the exact sequence
\begin{equation}
0\to \mathcal{I}_X/\mathcal{I}_X^2\stackrel{d}{\longrightarrow}\Omega^1_Y|_X\to \Omega^1_X\to 0.
\end{equation}
Letting $\pi:Y\to C$ denote the projection, we can rewrite this as
\begin{equation}
0\to \left(\pi^*(K_C^{-2})\right)|_X\stackrel{d}{\longrightarrow}
\left(\pi^*(K_C)\oplus\pi^*(L_1^*)\oplus \pi^*(L_2^*)\oplus \pi^*(K_C^*)\right)|_X\to
\Omega^1_X\to 0.
\label{eq:cotangent}
\end{equation}
The components of $d$ are given by the components of $df=x\,dy\,+y\,dx+2z\,dz$,
i.e.\ $(0,y,x,2z)$.  So we compute $\underline{Ext}^1_X$ by  dualizing $d$
\begin{equation}
  \label{eq:ext1}
\left(\pi^*(K_C^*)\oplus\pi^*(L_1)\oplus \pi^*(L_2)\oplus \pi^*(K_C)\right)|_X
\stackrel
{d^{\mathrm{t}}}
{\longrightarrow}
\left(\pi^*(K_C^{2})\right)|_X\to
\underline{Ext}^1_X(\Omega^1_X,\mathcal{O}_X)\to 0,
\end{equation}
i.e.\ $\underline{Ext}^1_X(\Omega^1_X,\mathcal{O}_X)$ is the cokernel of
$d^{\mathrm{t}}$.  But the map $d^{\mathrm{t}}$ imposes $x=y=z=0$ on the cokernel,
in other words, just restricts $\left(\pi^*(K_C^{2})\right)|_X$ to $C$.
Thus $\underline{Ext}^1_X(\Omega^1_X,\mathcal{O}_X)$
is just $K_C^2$, or more formally, $i_*(\mathcal{O}_C(K_C^2)$,  where $i:C\hookrightarrow
X$ is the inclusion.  So the local deformations are simply given by $H^0(C,K_C^2)$,
the base of the $SU(2)$ Hitchin system.

\bigskip
Next, we look at the $A_n$ case, with local model $xy=z^{N+1}$ in the total
space of $L_1\oplus L_2\oplus K_C$, where $L_1\otimes L_2\simeq K_C^{N+1}$.
Then (\ref{eq:cotangent}) is replaced by
\begin{equation}
0\to \left(\pi^*(K_C^{-{N+1}})\right)|_X\stackrel{d}{\longrightarrow}
\left(\pi^*(K_C)\oplus\pi^*(L_1^*)\oplus \pi^*(L_2^*)\oplus \pi^*(K_C^*)\right)|_X\to
\Omega^1_X\to 0.
\label{eq:cotangentn}
\end{equation}
with $d$ in components given by
$(0,y,x,(N+1)z^N)$.  Similarly, $\mathrm{Ext}^1_X(\Omega^1_X,\mathcal{O}_X)$
is now 
\begin{equation}
  \label{eq:ext1n}
\left(\pi^*(K_C^*)\oplus\pi^*(L_1)\oplus \pi^*(L_2)\oplus \pi^*(K_C)\right)|_X
\stackrel
{d^{\mathrm{t}}}
{\longrightarrow}
\left(\pi^*(K_C^{N+1})\right)|_X\to
\underline{Ext}^1_X(\Omega^1_X,\mathcal{O}_X)\to 0,
\end{equation}
with the new $d$.  So
$\underline{Ext}^1_X(\Omega^1_X,\mathcal{O}_X)$ is the restriction of
$\pi^*(K_C^{N+1})$ to the infinitesimal
neighborhood $C_{N-1}$ of $C$ defined by $x=y=z^N=0$, an invertible sheaf on
$C_{N-1}$.
\footnote{In \cite{Anderson:2013rka}, this was described
less precisely as a vector bundle on $C$.}

Let $\mathcal{I}\subset\mathcal{O}_{C_{N-1}}$
be the ideal of $C$ in $C_{N-1}$, locally generated by $z$.  We get a filtration of $\mathcal{O}_{C_{N-1}}$
\begin{equation}
  \mathcal{O}_{C_{N-1}}=\mathcal{I}^0\supset \mathcal{I}^1\supset\cdots\supset
\mathcal{\cI}^{N-1}\supset\mathcal{I}^N=0
\end{equation}
and a corresponding filtration of $\pi^*(K_C^{N+1})|_{C_{N-1}}$
\begin{equation}
  \pi^*(K_C^{N+1})|_{C_{N-1}}\supset \mathcal{I}^1\pi^*(K_C^{N+1})|_{C_{N-1}}
\supset\cdots\supset
\mathcal{\cI}^{N-1}\pi^*(K_C^{N+1})|_{C_{N-1}}\supset0.
\label{eq:k2filt}
\end{equation}
Now $\mathcal{I}/\mathcal{I}^2$ is the conormal bundle of $C$, isomorphic
to $K_C^*$.  It follows that $\mathcal{I}^k/\mathcal{I}^{k+1}\simeq K_C^{-k}$
for $k\le N-1$.  We can therefore break up (\ref{eq:k2filt}) into
short exact sequences
\begin{equation}
  \label{eq:filteredses}
  0\to\cI^{k+1}\pi^*(K_C^{N+1})|_{C_{N-1}}\to
\cI^{k}\pi^*(K_C^{N+1})|_{C_{N-1}}\to i_*(\mathcal{O}_C(K_C^{N+1-k}))\to 0,
\end{equation}
where by convention we put $\cI^0\pi^*(K_C^{N+1})=\pi^*(K_C^{N+1})$.
Observe that (\ref{eq:filteredses}) remains exact on
global sections
\begin{equation}
  \label{eq:filteredglobal}
  0\to H^0(\cI^{k+1}\pi^*(K_C^{N+1})|_{C_{N-1}})\to
H^0(\cI^{k}\pi^*(K_C^{N+1})|_{C_{N-1}})\to H^0(C,K_C^{N+1-k})\to 0,
\end{equation}
so that $H^0(X,\mathcal{T}^1)$ is filtered by the
$H^0(\cI^{k}\pi^*(K_C^{N+1})|_{C_{N-1}})$ with graded quotients
$H^0(C,K_C^{N+1-k})$,
and is thus (noncanonically) isomorphic to the Hitchin base $\oplus_{k=2}^{N+1}
H^0(C,K_C^k)$.

However, to match with gauge theory, we need a canonical isomorphism.  For this,
recall that $\pi^*(K_C)$ has a canonical $K_C$-valued section which we call $\lambda$,
familiar from the Hitchin system.  Using $\lambda$, we have a map
\begin{equation}
\bigoplus_{k=2}^{N+1}H^0(C,K_C^k)\stackrel{\phi}{\to}
H^0(C_{N-1},\pi^*(K_C^{N+1})),\qquad
(\omega_2,\ldots,\omega_{N+1})\mapsto
\left(\sum_{k=2}^{N+1}\pi^*\omega_k\lambda^{N+1-k}\right)|_{C_{N-1}}.
\label{eq:filt2sum}
\end{equation}
Considering successive quotients and taking the previous discussion into
consideration, we see immediately that the canonical map $\phi$ is
an isomorphism, thereby identifying $H^0(X,\mathcal{T}^1)$ with the
base of the $SU(N)$ Hitchin system in the local case.  Compare to \cite{Diaconescu:2006ry}.

We now allow the $A_{N-1}$ singularity to enhance to $A_N$ at isolated points $p_i$.
In F-theory, this corresponds to the transverse intersection of an $I_1$ curve with the
$I_N$ locus $C$.  This is the situation considered in \cite{Anderson:2013rka},
which we now review.

Let $D=\sum_{i=1}^d p_i$ be the corresponding divisor.  Our local model is
\begin{equation}
  xy=z^{N+1}+wz^N,
\end{equation}
where $w\in H^0(C,\mathcal{O}(D))$ is a section vanishing precisely at the
$p_i$.

In \cite{Anderson:2013rka},
it was shown that $\mathcal{T}^1$ has a torsion subsheaf $\mathrm{Tors}
(\mathcal{T}^1)$, a skyscraper sheaf supported on $D$, one-dimensional over
each $p_i$.  Then $\mathcal{T}^1$ is annihilated by the partial derivatives
$x,y,(n+1)z^N+Nwz^{N-1},z^N$, and the torsion class is generated by $z^{N-1}$.
So $\mathcal{T}^1/\mathrm{Tors}(\mathcal{T}^1)$ is annihilated by
$x,y,z^{N}$.  In other words, if
we let $C_{N-1}$ denote the non-reduced curve
with equation $x=y=z^N=0$ as in the $A_{N-1}$ case, then $\mathcal{T}^1/\mathrm{Tors}(\mathcal{T}^1)$ is the locally free sheaf associated to a
line bundle on $C_{N-1}$.

We have the short exact sequence
\begin{equation}
  0\to \mathrm{Tors}(\mathcal{T}^1)\to \mathcal{T}^1\to
\mathcal{T}^1/(\mathrm{Tors}(\mathcal{T}^1))\to 0.
\label{eq:t1ses}
\end{equation}
Since $H^1(\mathrm{Tors}(\mathcal{T}^1))=0$, (\ref{eq:t1ses}) remains exact
on global sections:
\begin{equation}
  0\to H^0(X,\mathrm{Tors}(\mathcal{T}^1))\to H^0(X,\mathcal{T}^1)\to
H^0(X,\mathcal{T}^1)/(\mathrm{Tors}(\mathcal{T}^1))\to 0.
\label{eq:h0t1ses}
\end{equation}

Concretely, the deformations are described as
\begin{equation}
  xy=z^{N+1}+wz^N+\sum_{j=1}^N \omega_jz^{N-j},
\label{eq:andefhiggs}
\end{equation}
with $\omega_j\in H^0(C,K_C^j(D))$.  The correspondence $\omega_j\leftrightarrow
\omega_j/w$ identifies global sections of $\mathcal{O}(K_C^j(D))$ with meromorphic sections
of $\mathcal{O}(K_C^j)$ with first order poles on $D$, and we will frequently make this
identification without comment.  Away from $D$, the
$u^{N-1}$ term can be eliminated by shifting $u$, so only the residues of
$\omega_1$ are true parameters.  The invariant way to say this is
\begin{equation}
  \mathrm{Tors}(\mathcal{T}^1)=\mathcal{O}(K_C(D))|_D,\qquad
  H^0(X,\mathrm{Tors}(\mathcal{T}^1))=H^0(D,\mathcal{O}(K_C(D))|_D).
\end{equation}
Comparing the above local description with the global description of
the pure $A_{N-1}$ case, we see that $\mathcal{T}^1/(\mathrm{Tors}(\mathcal{T}^1)$ is the invertible sheaf of sections of the line bundle
$\pi^*(K_C^N(D))|_{C_{N-1}}$ on $C_{N-1}$.  We
filter $\mathcal{T}^1/\mathrm{Tors}(\mathcal{T}^1)$ using powers of
$\mathcal{I}$ as before, use $\mathcal{I}/\mathcal{I}^2\simeq\mathcal{O}(K_C^*)$,
then finally use the spectral cover description as in (\ref{eq:filt2sum})
to canonically identify
\begin{equation}
  H^0(C,\mathcal{T}^1/(\mathrm{Tors}(\mathcal{T}^1)))\simeq
\bigoplus_{j=2}^N H^0(C,K_C^j(D)),
\end{equation}
the base of a parabolic Hitchin system.

We now identify these moduli with the Higgs branch of an $SU(N)$ gauge
theory.  For now, we
content ourselves
with describing the system in holomorphic gauge, ignoring the gauge field.
Stability will guarantee that a gauge field can be found which satisfies the D-term
constraints after going to unitary gauge.

We consider a Higgs field $\Phi$, and hypermultiplets $\psi_i \oplus \psi^c_i$
localized  at $p_i$, with $\psi_i$ in the $\mathbf{N}$ representation
and $\psi_i^c$ in the $\overline{\mathbf{N}}$.
Note that the $N\times N$ matrix $\psi_i\otimes \psi_i^c $ has rank less than
or equal to 1.  The rank can only be zero if either $\psi_i$ or $\psi_i^c$ is zero.  Otherwise, the rank is 1 and $\psi_i\otimes \psi_i^c$ is either a projection operator onto a one-dimensional
subspace, or nilpotent.  These two cases are distinguished by the non-vanishing
or vanishing of $\mathrm{Tr}(\psi_i\otimes \psi_i^c)$.  This trace appears in
the identity
\begin{equation}
  \label{eq:pairing}
\langle \langle \psi_j , \psi_j^c \rangle \rangle=\psi_i\otimes \psi_i^c-\frac1{N}  \mathrm{Tr}(\psi_i\otimes \psi_i^c)I_N,
\end{equation}
where $I_N$ is the $N\times N$ identity matrix.

Intrinsically we can identify $\mathrm{Tr}(\psi_i\otimes \psi_i^c)$
with the residue of a meromorphic 1-form at $p_i$,  identifying it
with the torsion deformations  $H^0(X,\mathrm{Tors}(\mathcal{T}^1))$.
If on the other hand $\mathrm{Tr}(\psi_i\otimes \psi_i^c)=0$, then
$\langle \langle \psi_j , \psi_j^c \rangle \rangle=\psi_i\otimes \psi_i^c$
itself has rank~1.  In the rank~0 case we similarly have
$\langle \langle \psi_j , \psi_j^c \rangle \rangle=\psi_i\otimes \psi_i^c\ (=0)$
and these cases can now be combined.

To describe the correspondence and allow non-trivial gauge bundles, we
let $\mathcal{P}$ denote a principal $G:=SU(N)$-bundle on $C$.
Then $\Phi\in \Gamma(\mathrm{ad}(\mathcal{P})\otimes K_C(D))$,
$\psi_i\in (\mathcal{P}\times_G\mathbf{N})_{p_i}$, and
$\psi_i^c\in (\mathcal{P}\times_G\overline{\mathbf{N}})|_{p_i}$.
The poles of $\Phi$ at the $p_i$ will be explained presently.

Then the equation for $\Phi$ is

\begin{equation}
 \bar\partial \Phi = \sum_j \delta_{p_j} \langle \langle \psi_j ,  \psi_j^c \rangle \rangle .
\label{eq:shiftdefect}
\end{equation}

Thus $\Phi$ can have first order poles at the $p_j$.

We can rephrase (\ref{eq:shiftdefect}) by saying that $\Phi$ is meromorphic,
with first order poles at the $p_i$ whose residues are determined by $\psi$
and $\psi^c$.  This perspective will be useful in generalizations.

We now match to geometry.  As mentioned above, the case $\mathrm{Tr}(\psi_i \otimes  \psi_i^c )\ne0$ corresponds to the torsion deformation at $p_i$.  Modding out by
the torsion deformations, we
suppose that $\mathrm{Tr}( \langle \langle \psi_i , \psi_i^c \rangle \rangle)=0$ for all $i$, so that
$\langle \langle \psi_i , \psi_i^c \rangle \rangle$ is nilpotent with square zero.

Multiplying by $w$ to clear the poles, we let $z\in\mathcal{O}(K_C)$ and consider the
characteristic polynomial
\begin{equation}
z^{n+1}+w \mathrm{det} (zI + \Phi),
\label{eq:shiftchar}
\end{equation}
which is of the form
\begin{equation}
z^{N+1}+wz^N+\sum_{j=2}^N \omega_jz^{N-j}.
\end{equation}
Note that all poles in the expansion of the determinant in (\ref{eq:shiftchar})
are simple since each $\langle \langle \psi_j , \psi_j^c \rangle \rangle$ has rank~1.
Hence, for $j\ge2$, we have $\omega_j\in H^0(K_C^j(D))$,
a precise match with the right hand side of  (\ref{eq:andefhiggs}).

\bigskip
Now we turn to an $A_{N-1}$-$A_{M-1}$ collision, with local equation
\begin{equation}
  xy=z^2+u^Nv^M.
\end{equation}

We let the two component curves be $C_1$ (corresponding to $u=0$) and $C_2$
(corresponding to $v=0$), intersecting transversally at the point $p$ corresponding to
$u=v=0$.  As before,
the sheaf $\mathcal{T}^1_X$ is a line bundle on the scheme $D$
locally defined by the partial derivatives, $x=y=z=u^{N-1}v^M=u^Nv^{M-1}=0$.
Then $D$ is given locally by $u^N=0$ along $C_1-p$ and by
$v^M=0$ along $C_2-p$.

Then $u^{N-1}v^{M-1}$
is a torsion class at $p$ as it is annihilated by each of $x,y,z,u,v$.  Write
$D_{\mathrm{CM}}$ for the reduction of $D$ modulo this torsion class. Thus
$D_{\mathrm{CM}}$ has the single local equation $u^{N-1}v^{M-1}=0$, showing that
$D_{\mathrm{CM}}$ is the union of the two (not necessarily reduced) components
${(C_1)}_{N-1}$
and ${(C_2)}_{M-1}$.  Then we have a short exact sequence
\begin{equation}
  0\to \mathrm{Tors}(\mathcal{T}^1_X)\to \mathcal{T}^1_X\to \mathcal{L}\to 0
\end{equation}
for some line bundle $\mathcal{L}$ on $D_{\mathrm{CM}}$.

Combining the previous discussion with the argument in the case of a defect
considered in \cite{Anderson:2013rka}, we have $\mathcal{L}|_{(C_1)_{N-1}}\simeq
\pi^*(K_{C_1}^N(Mp))$ and $\mathcal{L}|_{(C_2)_{M-1}}\simeq
\pi^*(K_{C_2}^M(Np))$, where $(C_1)_{N-1}$ is the thickening of $C_1$ defined by
$u^{N-1}=0$, and similarly $(C_2)_{M-1}$ is the thickening of $C_2$ defined by
$v^{M-1}=0$.

By restriction of $\mathcal{L}$ to the components we get
an exact sequence
\begin{equation}
  0\to \mathcal{L}\to \mathcal{L}|_{{(C_1)}_{N-1}}\oplus \mathcal{L}|_{{(C_2)}_{M-1}}
\to \mathcal{L}|_{u^{N-1}=v^{M-1}=0}\to 0,
\end{equation}
where the term  $\mathcal{L}|_{u^{N-1}=v^{M-1}=0}$ enforces an identification of sections
of $\mathcal{L}|_{{(C_1)}_{N-1}}$ and $\mathcal{L}|_{{(C_2)}_{M-1}}$ necessary to get a section
of $\mathcal{L}$.
This gives an exact sequence
\begin{equation}
  0\to H^0(\mathcal{L})\to H^0(\mathcal{L}|_{{(C_1)}_{N-1}})\oplus H^0(\mathcal{L}|_{{(C_2)}_{M-1}})
\to H^0(\mathcal{L}|_{u^{N-1}=v^{M-1}=0})\to 0,
\label{eq:lses}
\end{equation}
where the map on the right is seen to be surjective, either by explicit
calculation or by showing $H^1(\mathcal{L})=0$.

Combining these calculations with the
calculation in the pure $A_N$ case, we deduce a canonical isomorphism
\begin{equation}
  H^0(\mathcal{L}|_{{(C_1)}_{N-1}})\simeq H^0(K_{C_1}^2(Mp))\oplus\ldots\oplus
H^0(K_{C_1}^N(Mp)),
\label{eq:lc1}
\end{equation}
and similarly for the restriction of $\mathcal{L}$ to $(C_2)_{M-1}$.

For simplicity, we assume that $C_1$ has genus $g_1\ge 2$ (otherwise we can  assume that there
are sufficiently many collisions to ensure a nontrivial Hitchin base).  Then
$H^0(\mathcal{L}|_{{(C_1)}_{N-1}})$ has dimension $(N^2-1)(g_1-1)+M(N-1)$.  We get a similar
result for the restriction to $C_2$.

Since $H^0(\mathcal{L}|_{u^{N-1}=v^{M-1}=0})$ has dimension $(M-1)(N-1)$, we get
$\dim H^0(\mathcal{L})=(N^2-1)(g_1-1)+(M^2-1)(g_2-1)+M(N-1)+N(M-1)-(M-1)(N-1)$, or
\begin{equation}
\dim H^0(\mathcal{L})=(N^2-1)(g_1-1)+(M^2-1)(g_2-1)+MN-1.
\end{equation}
But $h^0(\mathcal{T}^1_X)=h^0(\mathcal{T}^1_X/(\mathrm{Tors}\mathcal{T}^1_X))
+h^0(\mathrm{Tors}\mathcal{T}^1_X)=h^0(\mathcal{L})+1$, and so
\begin{equation}
  h^0(\mathcal{T}^1_X)=(N^2-1)(g_1-1)+(M^2-1)(g_2-1)+MN.
\end{equation}
This is precisely the dimension of the Higgs branch of  an $SU(N)\times SU(M)$
gauge theory with $g_1$ $SU(N)$ adjoints, $g_2$ $SU(M)$ adjoints, and a bifundamental.

\subsection{Unfolding $A_2$}

Let us illustrate some of these considerations in the special case where we unfold an $A_2$ singularity, i.e. $N = 3$.
We now  suppose that the $A_1$ singularity is enhanced  to $A_2$ at a divisor $D$ in $C$,
where the points in $D$ can have multiplicity greater than 1.  In F-theory, this situation
arises from a collision between an $I_2$ and $I_1$ divisor where the components meet at
$D$, including multiplicities from tangencies.

For computing $\mathcal{T}^1$ we start with a
local model and then sort out how these fit together globally.  Near a point $p$ of
multiplicity $m$, we can take the local model
\begin{equation}
xy+z^3+t^mz^2=0.
\label{eq:orderm}
\end{equation}

Computing partial
derivatives, the sheaf $\mathcal{T}^1$
is locally just
$\mathcal{O}_X/(y,x,3z^2+2t^mz,t^{m-1}z^2)$.  But the element $z\in\mathrm{T}^1_X$
is a torsion class supported at $p$ since it is annihilated by  $x,y,z^2$, and
$t^{2m-1}$, which jointly vanish only at $p$.  In addition, $z$ is annihilated by
$zt^{m-1}$.  Hence the torsion subsheaf of $\mathcal{T}_X^1$ at $p$ is
isomorphic to $\mathcal{O}/(x,y,z^2,zt^{m-1},t^{2m-1})$, which has dimension
$3m-2$.  The quotient of $\mathcal{T}^1_X$ by the torsion is clearly
$\mathcal{O}/(x,y,z)$, a line bundle on $C$.
Globally this is
\begin{equation}
  \label{eq:torsm}
  0\to \mathrm{Tors}(\mathcal{T}^1_X)\to \mathcal{T}^1_X\to \mathcal{O}_C(K_C^2(D))
\to 0,
\end{equation}
where the line bundle at the end is identified globally exactly as in
\cite{Anderson:2013rka}.  If $D=\sum m_ip_i$, then
$\dim H^0(\mathrm{Tors}(\mathcal{T}^1_X))=\sum_i(3m_i-2)$.

Now, these additional moduli may appear to be inconsistent with the
rules for F-theory.  We reconcile these two viewpoints by showing
that these local deformations are not in the kernel of the map $\delta$
from (\ref{eq:localtoglobal}), so do not extend to global deformations in
F-theory.

We content ourselves with considering the case of simple enhancements in F-theory.\footnote{We thank W. Taylor for suggesting that we do this calculation.} Consider the F-theory model with $f$ and $g$ given by
\begin{equation}
  \label{eq:fgi2}
  f=-2h_{4+2n}^2+uf_{8+3n}+u^2f_{8+2n}+O(u^3),\qquad
g=3h_{4+2n}^3-u f_{8+3n}h_{4+2n}+u^2g_{12+4n}+O(u^3),
\end{equation}
generically an $I_2$ singularity.
From the leading order behavior of the discriminant
\begin{equation}
  \label{eq:di2}
u^2(-9f_{8+3n}^{2}h_{4+2n}^2+108 h_{4+2n}^3g_{12+4n} +108h_{4+2n}^{4}f_{8+2n})+O(u^3),
\end{equation}
we see that we have enhancements at the zeros of
\begin{equation}
-9f_{8+3n}^{2}+108 h_{4+2n}g_{12+4n} +108h_{4+2n}^{2}f_{8+2n},
\end{equation}
a set of $16+6n$ points, generically distinct.  We also have antisymmetric
matter at the zeros of $h_{4+2n}$, but these are singlets and can be
ignored.

We can deform away from this F-theory geometry by relaxing the constraints
\begin{equation}
  f_{8+4n}=-3h_{4+2n}^2,\ g_{12+6n}=2h_{4+2n}^3,\ g_{12+5n}=-f_{8+3n}h_{4+2n}.
\end{equation}
We study the map from $(f_{8+4n},g_{12+6n},g_{12+5n})$ to the space
$\mathrm{Ext}^1_X(\Omega^1_X,\mathcal{O}_X)$ of first order deformation,
and then compose with the map $\mathrm{Ext}^1_X(\Omega^1_X,\mathcal{O}_X)\to
H^0(X,\mathcal{T}^1_X)$ from (\ref{eq:localtoglobal}) by restricting
attention to a neighborhood of the singularity.

We shift coordinates to the singularity along $u=0$ by the change of
coordinates $x\mapsto X+h_{4+2n}$.  Then the Weierstrass equation becomes
\begin{equation}
  y^2=X^3+3h_{4+2n}X^2+f_{8+3n}uX+u^2(g_{12+4n}+f_{8+2n}h_{4+2n})+O(u^3).
\end{equation}
The singularity is located at $y=X=u=0$, and is visibly an $A_1$ generically.
The singularity is enhanced at that points where the discriminant
$f_{8+3n}^2-12h_{4+2n}(g_{12+4n}+f_{8+2n}h_{4+2n})$ of the
quadratic form $3h_{4+2n}X^2+f_{8+3n}uX+u^2(g_{12+4n}+f_{8+2n}h_{4+2n})$ vanishes.
Of course this is precisely the divisor $D$ of $16+6n$ points discussed above.

The terms involving $(f_{8+4n},g_{12+6n},g_{12+5n})$ change the Weierstrass
equation by
\begin{equation}
  f_{8+4n}(X+h_{4+2n})+g_{12+6n}+g_{12+5n}u.
\end{equation}
Comparing with the discussion earlier in this section, we see that the terms
$f_{8+4n}X+g_{12+5n}u$ which vanish on the singularity must correspond to
torsion deformations, while the other terms $f_{8+4n}h_{4+2n}+g_{12+6n}$
correspond to
$H^0(\mathcal{T}^1/\mathrm{Tors}(\mathcal{T}^1))\simeq H^0(C,K_C^2(D))$.  However,
since $D$ has degree $16+6n$ and $C\simeq\mathbb{P}^1$, we have
$\mathcal{O}(K_C^2(D))\simeq\mathcal{O}_{\mathbb{P}^1}(12+6n)$.  This is as it must
be, since $f_{8+4n}h_{4+2n}+g_{12+6n}$ has degree $12+6n$.

To probe $\mathrm{Tors}(\mathcal{T}^1))$, we require
$f_{8+4n}h_{4+2n}+g_{12+6n}=0$, i.e.\ $g_{12+6n}=-f_{8+4n}h_{4+2n}$.  Then the free
moduli which map to $\mathrm{Tors}(\mathcal{T}^1)$ correspond to the
deformations
\begin{equation}
f_{8+4n}X+g_{12+5n}u.
\end{equation}
Let's study the kernel of the map sending the deformation
$f_{8+4n}X+g_{12+5n}u$ to $H^0(X,\mathcal{T}^1)$.  By the calculation of
$\mathcal{T}^1$, the kernel is generated by the partial derivatives of
$3h_{4+2n}X^2+f_{8+3n}uX+u^2(g_{12+4n}+f_{8+2n}h_{4+2n})$, i.e.\
\begin{equation}
  6h_{4+2n}X+f_{8+3n}u,\qquad f_{8+3n}X+2u(g_{12+4n}+f_{8+2n}h_{4+2n}).
\end{equation}
So the deformations which map to zero are generated by these two polynomials,
and we conclude by comparing degrees that the kernel is given by
\begin{equation}
  f_{8+4n}X+g_{12+5n}u=
p_{4+2n}\left(6h_{4+2n}X+f_{8+3n}u\right)+q_{n}\left(f_{8+3n}X+2u\left(
g_{12+4n}+f_{8+2n}h_{4+2n}\right)\right)
\label{eq:genker}
\end{equation}
for arbitrary polynomials $p_{4+2n}$ and $q_n$ of respective degrees $4+2n$
and $n$.  Furthermore, for generic moduli there are no redundancies in
the expression (\ref{eq:genker}).  So the dimension of the kernel is
\begin{equation}
  (2n+5)+(n+1)=3n+6.
\end{equation}
Since the space of the expressions $f_{8+4n}X+g_{12+5n}u$ has dimension
$(4n+9)+(5n+13)=9n+22$, we conclude that the space of torsion deformations
realized in our F-theory geometry is $(9n+22)-(3n+6)=6n+16$.
Since the space of torsion deformations is also $6n+16$, the degree of $D$,
we conclude that for generic moduli {\em all torsion deformations are
realized in F-theory\/}.

The situation is different for non-generic deformations.  Suppose that
we specialize to a situation where $m$ of the $16+6n$ enhancement points
coalesce, leaving $16+6n-m$ ordinary enhancement points.
By the calculation above, we get a $(16+6n-m)+(3m-2)=14+6n+2m$ dimensional
space of torsion deformations.  But the above calculation of the kernel
remains unchanged.  The only difference is that we have $m-1$ additional moduli
for deforming a pole of order $m$ to $m$ isolated poles.  So the
space of torsion deformations realized in our F-theory geometry is at most
$(6n+16)+(m-1)=6n+m+15$.  Since $6n+m+15 < 6n+2m+14$ for $m>1$, we conclude
that not all of the local torsion deformations associated with a pole of order
$m$ can be realized in a compact F-theory geometry.

It would be interesting to understand the matter associated with the
singularity (\ref{eq:orderm}) in type IIA, and what in the
gauge theory might prevent some of these putative deformations
from being realized in a compact F-theory geometry.

\newpage

\bibliographystyle{utphys}
\bibliography{WildTbranesFINAL}
\end{document}